\documentclass[twocolumn,english,prb,preprintnumbers,amsmath,amssymb,superscriptaddress]{revtex4}
\usepackage[latin9]{inputenc}
\usepackage{amsmath}
\usepackage{graphicx}

\makeatletter
\@ifundefined{textcolor}{}
{%
 \definecolor{BLACK}{gray}{0}
 \definecolor{WHITE}{gray}{1}
 \definecolor{RED}{rgb}{1,0,0}
 \definecolor{GREEN}{rgb}{0,1,0}
 \definecolor{BLUE}{rgb}{0,0,1}
 \definecolor{CYAN}{cmyk}{1,0,0,0}
 \definecolor{MAGENTA}{cmyk}{0,1,0,0}
 \definecolor{YELLOW}{cmyk}{0,0,1,0}
}

\usepackage{dcolumn}
\usepackage{bm}
\usepackage{color}
\usepackage[final]{pdfpages}

\makeatother

\usepackage{babel}

\begin{document}

\preprint{preprint(\today)}

\title{Time-reversal symmetry-breaking charge order in a kagome superconductor}

\author{C.~Mielke III}
\thanks{These authors contributed equally to the paper.}
\affiliation{Laboratory for Muon Spin Spectroscopy, Paul Scherrer Institute, CH-5232 Villigen PSI, Switzerland}
\affiliation{Physik-Institut, Universit\"{a}t Z\"{u}rich, Winterthurerstrasse 190, CH-8057 Z\"{u}rich, Switzerland}

\author{D.~Das}
\thanks{These authors contributed equally to the paper.}
\affiliation{Laboratory for Muon Spin Spectroscopy, Paul Scherrer Institute, CH-5232 Villigen PSI, Switzerland}

\author{J.-X.~Yin}
\thanks{These authors contributed equally to the paper.}
\affiliation{Laboratory for Topological Quantum Matter and Advanced Spectroscopy (B7), Department of Physics,
Princeton University, Princeton, New Jersey 08544, USA}

\author{H.~Liu}
\thanks{These authors contributed equally to the paper.}
\affiliation{Beijing National Laboratory for Condensed Matter Physics, Institute of Physics, Chinese Academy of Sciences, Beijing 100190, China}
\affiliation{School of Physical Sciences, University of Chinese Academy of Sciences, Beijing 100190, China}

\author{R.~Gupta}
\affiliation{Laboratory for Muon Spin Spectroscopy, Paul Scherrer Institute, CH-5232 Villigen PSI, Switzerland}

\author{Y.-X.~Jiang}
\affiliation{Laboratory for Topological Quantum Matter and Advanced Spectroscopy (B7), Department of Physics,
Princeton University, Princeton, New Jersey 08544, USA}

\author{M.~Medarde}
\affiliation{Laboratory for Multiscale Materials Experiments, Paul Scherrer Institut, CH-5232 Villigen PSI, Switzerland}

\author{X. Wu}
\affiliation{Max-Planck-Institut fur Festkorperforschung, Heisenbergstrasse 1, D-70569 Stuttgart, Germany}

\author{H.C.~Lei}
\affiliation{Department of Physics and Beijing Key Laboratory of Opto-electronic Functional Materials and Micro-nano Devices,
Renmin University of China, Beijing 100872, China}

\author{J.J. Chang}
\affiliation{Physik-Institut, Universit\"{a}t Z\"{u}rich, Winterthurerstrasse 190, CH-8057 Z\"{u}rich, Switzerland}

\author{P.~Dai}
\affiliation{Department of Physics and Astronomy, Rice University, Houston, Texas 77005, USA}

\author{Q.~Si}
\affiliation{Department of Physics and Astronomy, Rice University, Houston, Texas 77005, USA}

\author{H.~Miao}
\affiliation{Material Science and Technology Division, Oak Ridge National Laboratory, Oak Ridge, Tennessee 37831, USA}

\author{R. Thomale}
\affiliation{Institut fur Theoretische Physik und Astrophysik, Universitat Wurzburg, 97074 Wurzburg, Germany}
\affiliation{Department of Physics, Indian Institute of Technology Madras, Chennai 600036, India}

\author{T. Neupert}
\affiliation{Physik-Institut, Universit\"{a}t Z\"{u}rich, Winterthurerstrasse 190, CH-8057 Z\"{u}rich, Switzerland}

\author{Y.~Shi}
\affiliation{Beijing National Laboratory for Condensed Matter Physics, Institute of Physics, Chinese Academy of Sciences, Beijing 100190, China}
\affiliation{School of Physical Sciences, University of Chinese Academy of Sciences, Beijing 100190, China}
\affiliation{Songshan Lake Materials Laboratory, Dongguan, Guangdong 523808, China}

\author{R.~Khasanov}
\affiliation{Laboratory for Muon Spin Spectroscopy, Paul Scherrer Institute, CH-5232 Villigen PSI, Switzerland}

\author{M.Z. Hasan}
\email{mzhasan@princeton.edu} 
\affiliation{Laboratory for Topological Quantum Matter and Advanced Spectroscopy (B7), Department of Physics,
Princeton University, Princeton, New Jersey 08544, USA}
\affiliation{Princeton Institute for the Science and Technology of Materials, Princeton University, Princeton, New Jersey 08540, USA}
\affiliation{Materials Sciences Division, Lawrence Berkeley National Laboratory, Berkeley, California 94720, USA}
\affiliation{Quantum Science Center, Oak Ridge, Tennessee 37831, USA}

\author{H.~Luetkens}
\affiliation{Laboratory for Muon Spin Spectroscopy, Paul Scherrer Institute, CH-5232
Villigen PSI, Switzerland}

\author{Z.~Guguchia}
\email{zurab.guguchia@psi.ch} 
\affiliation{Laboratory for Muon Spin Spectroscopy, Paul Scherrer Institute, CH-5232 Villigen PSI, Switzerland}


\maketitle

\textbf{The kagome lattice\cite{Syozi}, the most prominent structural motif in quantum physics, benefits from inherent nontrivial geometry to host diverse quantum phases, ranging from spin-liquid phases, topological matter to intertwined orders \cite{Barz,ZHou,GuguchiaCSS,JXYin3,CMielke,TbNature,Pershoguba}, and most rarely unconventional superconductivity \cite{CMielke,BOrtiz2}. Recently, charge sensitive probes have suggested that the kagome superconductors $A$V$_{3}$Sb$_{5}$ ($A$ = K, Rb, Cs) \cite{BOrtiz2,BOrtiz3,QYin} exhibit unconventional chiral charge order \cite{YJiang,NShumiya,Wang2021,XFeng,MDenner,PLin,XWu,Qimiao}, which is analogous to the long-sought-after quantum order in the Haldane model \cite{Haldane} or Varma model \cite{Varma}. However, direct evidence for the time-reversal symmetry-breaking of the charge order remains elusive. Here we utilize muon spin relaxation to probe the kagome charge order and superconductivity in KV$_{3}$Sb$_{5}$. We observe a striking enhancement of the internal field width sensed by the muon ensemble, which takes place just below the charge ordering temperature and persists into the superconducting state. Remarkably, the muon spin relaxation rate below the charge ordering temperature is substantially enhanced by applying an external magnetic field. We further show the multigap nature of superconductivity in KV$_{3}$Sb$_{5}$ and that the $T_{\rm c}$/$\lambda_{ab}^{-2}$ ratio is comparable to those of unconventional high-temperature superconductors. Our results point to time-reversal symmetry breaking charge order intertwining with unconventional superconductivity in the correlated kagome lattice.} \\

\begin{figure*}[t!]
\centering
\includegraphics[width=0.95\linewidth]{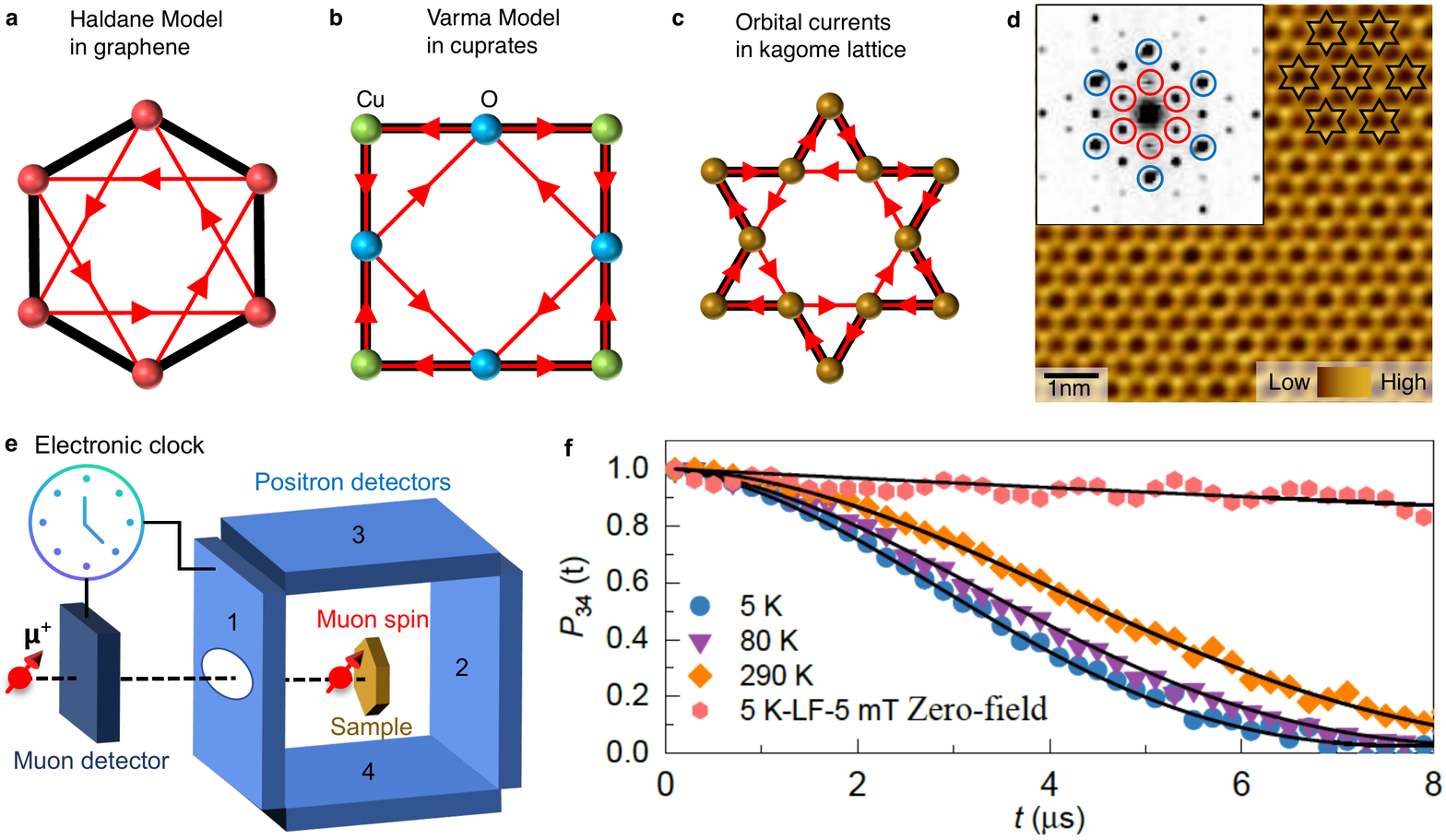}

\vspace{-1.5cm}

\includegraphics[width=0.95\linewidth]{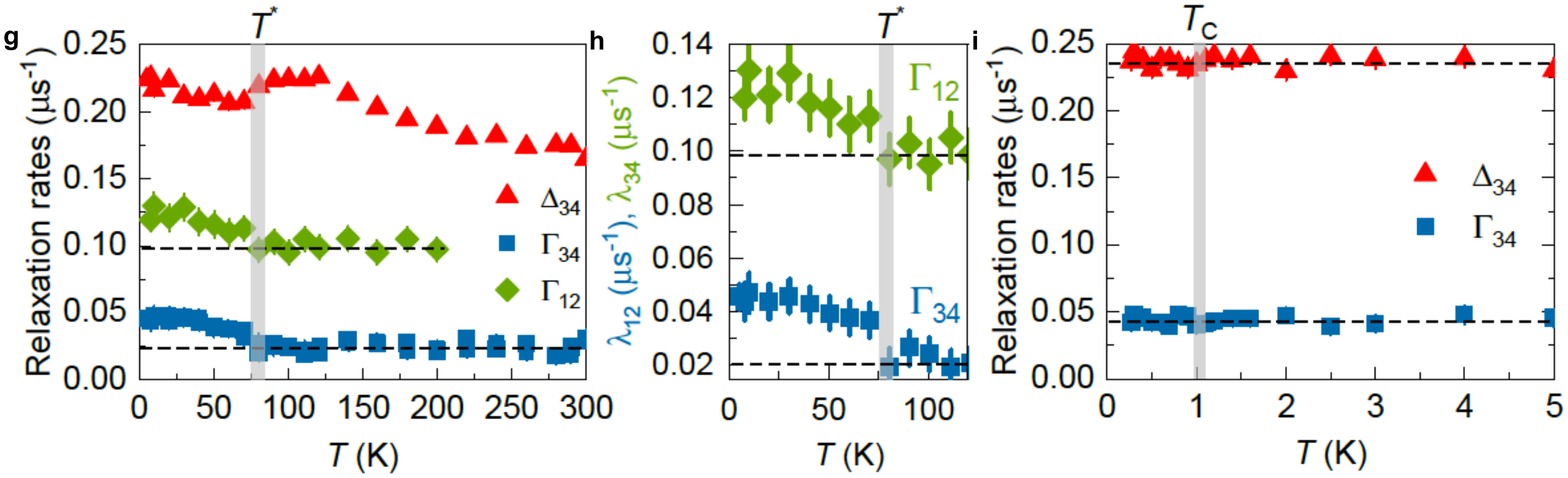}
\vspace{-6.5cm}
\caption{\textbf{Indication of time-reversal symmetry-breaking of the charge order in KV$_{3}$Sb$_{5}$.} 
$\bf{a,}$ Orbital currents (red arrows) proposed in a honeycomb lattice. $\bf{b,}$ Orbital currents (red arrows) proposed in the CuO$_{2}$ lattice of cuprates. $\bf{c,}$ Schematic of the orbital currents (red arrows) in the kagome lattice. $\bf{d,}$ Scanning tunneling microscopy of the Sb surface showing 2${\times}$2 charge order as illustrated by black lines. The inset is the Fourier transform of this image, shown lattice Bragg peaks marked by blue circles and 2${\times}$2 vector peaks marked by red circles. The three pairs of 2${\times}$2 vector peaks feature different intensities, denoting a chirality of the charge order. $\bf{e,}$ A schematic overview of the experimental setup. Spin polarized muons with spin $S_{\mu}$, forming $60^{\circ}$ with respect to the $c$-axis of the crystal, are implanted in the sample. The sample was surrounded by four detectors: Forward (1), Backward (2), Up (3), and Down (4). An electronic clock is started at the time the muon passes the muon detector and is stopped as soon as the decay positron is detected in the positron detectors. $\bf{f,}$ The ZF ${\mu}$SR time spectra for KV$_{3}$Sb$_{5}$, obtained at different temperatures, all above the superconducting transition temperature $T_{\rm c}$. The solid black curves in panel a represent fits to the recorded time spectra, using the Eq. 1. Error bars are the standard error of the mean (s.e.m.) in about 10$^{6}$ events. $\bf{g,}$ The temperature dependences of the relaxation rates ${\Delta}$ and ${\Gamma}$, obtained in a wide temperature range. $\bf{h,}$ The temperature dependence of ${\Gamma}$ from two sets of detectors across the charge ordering temperature $T^{*}$ ${\simeq}$ 80 K. $\bf{i,}$ Temperature dependences of the muon spin relaxation rates ${\Delta}$ and ${\Gamma}$, which can be related to the nuclear and electronic system respectively, in the temperature range across $T_{\rm c}$. The error bars represent the standard deviation of the fit parameters.}
\label{fig1}
\end{figure*}

\textbf{Introduction}   
  
 The observation of orbital currents is a long standing motive in both topological and correlated quantum matter. They have been suggested to produce the quantum anomalous Hall effect when interacting with Dirac fermions in a honeycomb lattice\cite{Haldane} (Fig.~1$\bf{a}$) and as the hidden phase of high-temperature cuprate superconductors\cite{Varma,Chakravarty} (Fig.~1$\bf{b}$). In both cases, orbital currents run through the lattice and break time-reversal symmetry. Recently, the tantalizing visualization of such exotic order has been reported \cite{YJiang,NShumiya,Wang2021} in the kagome superconductor $A$V$_{3}$Sb$_{5}$ ($A$ = K, Rb, Cs) (Fig.~1$\bf{c}$). Scanning tunneling microscopy observes a chiral 2${\times}$2 charge order (Fig.~1$\bf{d}$) with an unusual magnetic field response. Theoretical analysis \cite{YJiang,NShumiya,Wang2021,XFeng,MDenner,PLin,XWu,Qimiao} also suggests that this chiral charge order can not only lead to a giant anomalous Hall effect \cite{SYang} but also be a precursor of unconventional superconductivity \cite{XWu}. However, the broken time-reversal symmetry nature of the charge order and its interplay with superconductivity has not been explicitly demonstrated by experiments.  
 
  To explore unconventional aspects of superconductivity and the possible time-reversal symmetry breaking nature of charge order and superconductivity in KV$_{3}$Sb$_{5}$, it is critical to measure the superconducting order parameter and weak internal fields of KV$_{3}$Sb$_{5}$ on the microscopic level. Thus, we concentrate on muon spin relaxation/rotation (${\mu}$SR) experiments  \cite{Sonier} of the normal state depolarization rate and the magnetic penetration depth $\lambda$ in KV$_{3}$Sb$_{5}$. Importantly, zero-field ${\mu}$SR (ZF-${\mu}$SR) has the ability to detect internal magnetic fields as small as 0.1~G without applying external magnetic fields, making it a highly valuable tool for probing spontaneous magnetic fields due to time-reversal symmetry breaking \cite{LukeTRS} within the superconducting and charge ordered states.\\
     
\textbf{Results and Discussion} \\     
\textbf{Magnetic Response Across Charge Order}\\
     
   While long-range magnetism has not been reported in KV$_{3}$Sb$_{5}$ \cite{EKenney}, zero-field ${\mu}$SR experiments have been carried out above and below $T_{{\rm c}}$ to search for any weak magnetism (static or slowly fluctuating). A schematic overview of the experimental setup with the muon spin forming $60^{\circ}$ with respect to the $c$-axis of the crystal is shown in the inset of Figure~1e. The sample was surrounded by four detectors: Forward (1), Backward (2), Up (3), and Down (4). Figure 1f displays the zero-field ${\mu}$SR spectra from detectors 3 \& 4 collected over a wide temperature range. We see that the muon spin relaxation shows a reasonable temperature dependence. Since the zero-field relaxation is decoupled by the application of a small external magnetic field applied longitudinal to the muon spin polarization, $B_{{\rm LF}}$~=~50~G (see Fig.~1$\bf{f}$), the relaxation is therefore due to spontaneous fields which are static on the microsecond timescale. The zero-field ${\mu}$SR spectra were fitted using the gaussian Kubo-Toyabe depolarization function \cite{Toyabe}, which reflects the field distribution at the muon site created by the nuclear moments of the sample, multiplied by an additional exponential exp(-$\Gamma t$) term (see also the methods section):

\begin{equation}
\begin{aligned}
P_{ZF}^{GKT}(t) =  \left(\frac{1}{3} + \frac{2}{3}(1 - \Delta^2t^2 ) \exp\Big[-\frac{\Delta^2t^2}{2}\Big]\right) \exp(-\Gamma t) \\ 
\end{aligned}
\end{equation}  
 where ${\Delta}$/${\gamma_{\mu}}$ is the width of the local field distribution due to the nuclear moments and ${\gamma_{\mu}}$/2${\pi}$ = 135.5~MHz/T is the muon gyromagnetic ratio. The observed deviation from a pure GKT behavior in paramagnetic systems is frequently observed in ${\mu}$SR measurements.  This can e.g. be due to a mixture of diluted and dense nuclear moments, the presence of electric field gradients or a contribution of electronic origin. A gaussian Kubo Toyabe shape is expected due to the presence of the dense system of nuclear moments with large values of nuclear spins ($I$ = 3/2 for $^{39}$K, $I$ = 7/2 for $^{51}$V, and $I$ = 5/2 for $^{121}$Sb) in KV$_{3}$Sb$_{5}$ and a high natural abundance. 
The relaxation in single crystals might also be not GKT-like due to the fact that the quantization axis for the nuclear moments depends on the electric field gradients \cite{Sonier2012}. Naturally this is also often responsible for an anisotropy of the nuclear relaxation. As this effect essentially averages out in polycrystalline samples, we would like to mention that we also observed the additional exponential term in the polycrystalline sample of KV$_{3}$Sb$_{5}$ (see the methods section) which indicates that this effect is probably not the dominant in our single crystal measurements. Our high field ${\mu}$SR results presented below however prove that there is indeed a strong contribution of electronic origin to the muon spin relaxation below the charge ordering temperature. Therefore, we conclude that ${\Gamma}$ in zero magnetic field also tracks the temperature dependence of the electronic contribution, but cannot exclude subtle effects due to changes in the electric field gradients in the charge ordered state. In Fig.~1$\bf{g}$, we see the temperature dependence of both the muon spin relaxation  ${\Delta}_{12}$ and ${\Gamma}$$_{12,34}$ over a broad temperature range from the base temperature to 300~K. There is a noteworthy increase immediately visible in the relaxation rates ${\Gamma}$$_{12}$ and ${\Gamma}$$_{34}$ upon lowering the temperature below the charge ordering temperature $T^{*}$, which is better visible in Fig.~1$\bf{h}$. This observation indicates the enhanced spread of internal fields sensed by the muon ensemble concurrent with the onset of charge ordering. The enhanced magnetic response that sets in with the charge order persists all the way down to the base temperature, and remains constant across the superconducting transition, as seen in Fig.~1$\bf{i}$. Increase of the internal field width visible from the ZF-${\mu}$SR relaxation rate corresponds to an anomaly seen also in the nuclear contribution to the relaxation rate ${\Delta}_{12}$; namely, a peak coinciding with the onset of the charge order, which decreases to a broad minimum before increasing again towards lower temperatures.    

\begin{figure*}[t!]
\includegraphics[width=1.0\linewidth]{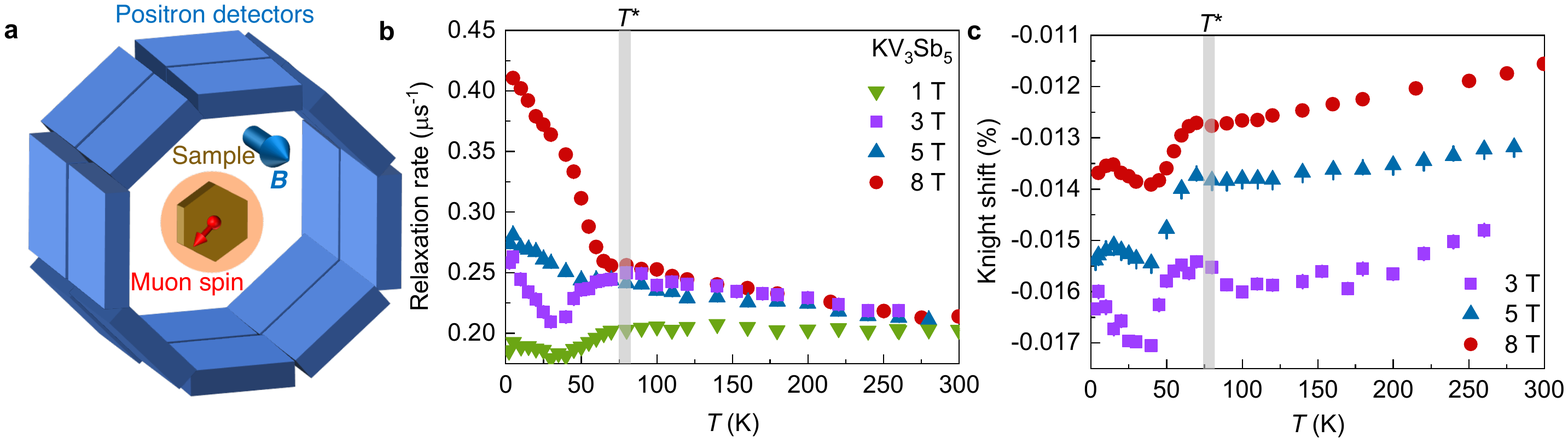}
\vspace{-7.7cm}
\caption{\textbf{Enhanced magnetic response of the charge order with applying external magnetic fields.} 
$\bf{a,}$ Schematic overview of the high field ${\mu}$SR experimental setup for the muon spin forming $90^{\circ}$ with respect to the $c$-axis of the crystal. The sample was surrounded by 2 times 8 positron detectors, arranged in rings. The specimen was mounted in a He gas-flow cryostat with the largest face perpendicular to the muon beam direction, along which the external field was applied. Behind the sample lies a veto counter (in orange) which rejects the muons that do not hit the sample. The temperature dependence of $\bf{b,}$ the muon spin relaxation rate and $\bf{c,}$ the Knight shift $K_{\rm exp}$ (local susceptibility) for KV$_{3}$Sb$_{5}$, measured under the $c$-axis magnetic fields of ${\mu}$$_{0}H$ = 1 T, 3 T, 5 T, and 8 T. The vertical grey lines mark the charge ordering temperature, determined from magnetization measurements (see the methods section). $K_{\rm exp}$ also shows a shallow minimum at around 30 K, followed by a small peak towards low temperatures. The error bars represent the standard deviation of the fit parameters.}
\label{fig3}
\end{figure*}

 The increase in the exponential relaxation below $T^{*}$ is estimated to be ${\simeq}$ 0.025~${\mu}$$s^{-1}$, which can be interpreted as a characteristic field strength ${\Gamma}$$_{12}$/${\gamma_{\mu}}$ ${\simeq}$ 0.3~G. We note that a similar value of internal magnetic field strength is reported in several time-reversal symmetry-breaking superconductors \cite{LukeTRS}. Dip-like temperature dependence of ${\Delta}_{12}$ is also reminiscent of the behavior observed in some multigap TRS broken superconductors (e.g. $La_{7}$$Ni_{3}$ \cite{Singh}) across $T_{{\rm c}}$. However, in the present case the ZF-${\mu}$SR results alone do not allow to conclude on the time-reversal symmetry-breaking  effect in KV$_{3}$Sb$_{5}$ below $T^{*}$. As said above, the onset of charge order might also alter the electric field gradient experienced by the nuclei and correspondingly the magnetic dipolar coupling of the muon to the nuclei \cite{Sonier2012}. This can induce a change in the nuclear dipole contribution to the zero-field ${\mu}$SR signal. In order to substantiate the above zero-field ${\mu}$SR results, systematic high field ${\mu}$SR \cite{Sedlak} experiments are essential (Fig. 2$\bf{a}$). In a high magnetic field, the direction of the applied field defines the quantization axis for the nuclear moments, so that the effect of the charge order on the electric field gradient at the nuclear sites is irrelevant. A non-monotonous behavior of the relaxation rate is clearly seen in the ${\mu}$SR data, measured in magnetic field of 1~T, applied parallel to the $c$-axis, as shown in Fig.~2$\bf{b}$. The data at 1~T looks similar to the temperature dependence of the zero-field nuclear rate ${\Delta}_{12}$, it seems to be dominated by the nuclear response. However, at higher fields such as 3~T, 5~T and 8~T, the rate not only shows a broad bump around $T^{*}$, but also shows a clear and stronger increase towards low temperatures within the charge ordered state, similar to the behavior observed for the relaxation rates ${\Gamma}$$_{12}$ and ${\Gamma}$$_{34}$ in zero-field. As the nuclear contribution to the relaxation cannot be enhanced by an external field, this indicates that the low-temperature relaxation rate in magnetic fields higher than 1~T is dominated by the electronic contribution. Remarkably, we find that absolute increase of the relaxation rate between  the onset of charge order $T^{*}$ and the base-T in 8~T is 0.15 ${\mu}$$s^{-1}$ which is a factor of six higher than the one 0.025 ${\mu}$$s^{-1}$ observed in zero-field. This shows a strong field-induced enhancement of the electronic response. Moreover, we find that the magnitude of the Knight shift (local magnetic susceptibility), defined as $K_{\rm exp}$ = ($B_{\rm int}$ $-$ $B_{\rm ext}$)/$B_{\rm ext}$ ($B_{\rm int}$ and $B_{\rm ext}$ are the internally measured and externally applied magnetic fields) and obtained in 3~T, 5~T and 8~T shows a sharp increase just below $T^{*}$, as shown in Figure~2$\bf{c}$. The change in local magnetic susceptibility across the charge order temperature $T^{*}$ agrees well with the change seen in the macroscopic susceptibility (see the methods section) and indicate the presence of the magnetic response in KV$_{3}$Sb$_{5}$ concurrent with the charge order. $K_{\rm exp}$ shows a shallow minimum near 30 K at 3 T, 5 T, and 8 T, which is also seen in macroscopic susceptibility. The minimum in $K_{\rm exp}$ is followed by a small peak towards low temperatures, which is absent in macroscopic susceptibility. At present, it is difficult to give a quantitative explanation on the precise origin of such a behavior. However, in connection to previous experimental results, one possibility is that the dip-like feature and the observed peak is related to the transition from isotropic charge order to a low temperature electronic nematic state \cite{Xiang2021,Zhao2021}, which breaks rotational symmetry. 
Electronic nematic transition within the charge ordered state was reported for the related system CsV$_{3}$Sb$_{5}$ from transport \cite{Xiang2021} and STM experiments \cite{Zhao2021}. Appearance of a nematic susceptibility would certainly influence the Knight shift as well as the muon spin relaxation rate. On the other hand, changes in the charge section will also modify a hyperfine contact field at the ${\mu}^{+}$ site and thus the local susceptibility. If so, the modified local susceptibility will be reflected in a breakdown of the proportionality of the ${\mu}^{+}$ Knight shift to the measured bulk susceptibility since in this case local susceptibility is different from macroscopic susceptibility. This may explain different temperature dependence of muon Knight shift and macroscopic susceptibility within charge ordered state.

\begin{figure*}[t!]
\centering
\includegraphics[width=1.0\linewidth]{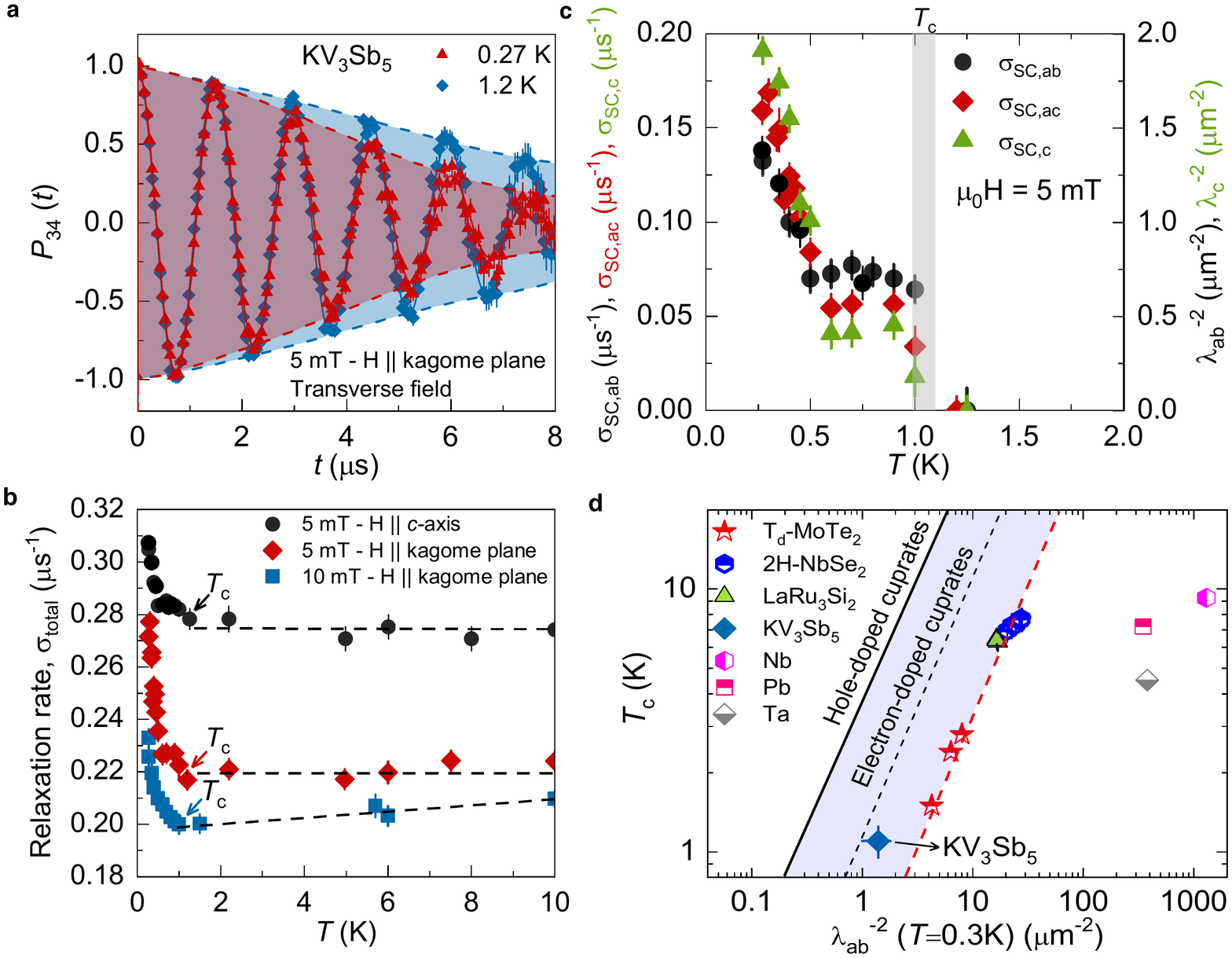}
\vspace{-1.2cm}
\caption{\textbf{Correlated kagome superconductivity.} 
$\bf{a,}$  The transverse field ${\mu}$SR spectra are obtained above and below $T_{\rm c}$ (after field cooling the sample from above $T_{\rm c}$). Error bars are the standard error of the mean (s.e.m.) in about 10$^{6}$ events. The error of each bin count n is given by the standard deviation (s.d.) of n. The errors of each bin in $A(t)$ are then calculated by s.e. propagation. The solid lines in panel a represent fits to the data by means of Eq.~3. The dashed lines are a guide to the eye. $\bf{b,}$  The temperature dependence of the total muon spin relaxation rate $\sigma$$_{\rm total}$ measured measured in the magnetic fields of 5 mT and 10 mT applied both parallel to the $c$-axis and parallel to the kagome plane. The dashed lines mark the average value of $\sigma$$_{\rm total}$ estimated from few data points above $T_{\rm c}$. 
$\bf{c,}$  The superconducting muon depolarization rates ${\sigma}_{SC,ab}$, ${\sigma}_{SC,ac}$, ${\sigma}_{SC,c}$ as well as the inverse squared magnetic penetration depth ${\lambda}_{ab}^{-2}$ and ${\lambda}_{c}^{-2}$ as a function of temperature, measured in 5 mT, applied parallel and perpendicular to the kagome plane. $\bf{d,}$ Plot of $T_{\rm c}$ versus the ${\lambda}_{ab}^{-2}(0)$ obtained from our ${\mu}$SR experiments in KV$_{3}$Sb$_{5}$. The dashed red line represents the relation obtained for kagome superconductor LaRu$_{3}$Si$_{2}$ as well as for the layered transition metal dichalcogenide superconductors $T_{d}$-MoTe$_{2}$ and 2H-NbSe$_{2}$ \cite{GuguchiaNbSe2}. The relation observed for underdoped cuprates is also shown (solid line for hole doping \cite{Uemura1} and the dashed black line for electron doping \cite{Luetkens,Shengelaya}). The points for various conventional Bardeen-Cooper-Schrieffer superconductors are also shown. The error bars represent the standard deviation of the fit parameters.}
\label{fig4}
\end{figure*}
 
The combination of ZF-${\mu}$SR and high-field ${\mu}$SR results show the enhanced internal field width below $T^{*}$, giving direct evidence for the time-reversal symmetry breaking fields in the kagome lattice. It is important to note that the increase of the relaxation rate arises from nearly the entire sample volume (see the methods section), indicating the bulk nature of the transition below $T^{*}$. Such observation is consistent with charge sensitive probe results that the magnetic field switching of the chiral charge order is observed in the impurity-free region \cite{YJiang}. Both the ${\mu}$SR and charge probe results attest to the intrinsic nature of the time-reversal breaking in the kagome lattice. One plausible phenomenological scenario is that the charge order has a complex chiral order parameter, which exhibits a phase difference between three sublattices of the kagome plane. The existence of phase difference, if not ${\pi}$, breaks time-reversal symmetry. Recent theoretical modeling of the charge ordering in the kagome lattice at van Hove filling and with extended Coulomb interactions (that is close to the condition of the $A$V$_3$Sb$_5$) also suggests that time-reversal symmetry-broken charge order with orbital currents is energetically favorable \cite{MDenner,PLin,XWu,Qimiao}. The orbital currents do not break translation symmetry beyond the 2${\times}$2 supercell of the charge order. In addition, at least according to the calculations \cite{MDenner}, the net flux in a 2${\times}$2 unit cell of the order is vanishingly small. Hence, there is extremely small net magnetic moment according to the theoretical modelling. The suggested orbital current was reported to be homogeneous on the lattice, however alternating in its flow, which would produce inhomogeneous fields at the muon site. Within this framework, muons may couple to the closed current orbits below $T^{*}$, leading to an enhanced internal field width sensed by the muon ensemble concurrent with the charge order. Thus, we conclude that the present results provide key evidence for a time-reversal symmetry broken charge order in KV$_{3}$Sb$_{5}$. However, we cannot determine the exact structure of orbital currents. Our data will inspire future experiments, particularly neutron polarization analysis, to potentially understand the precise order of orbital currents in KV$_{3}$Sb$_{5}$. The current results indicate that the magnetic and charge channels of KV$_{3}$Sb$_{5}$ appear to be strongly intertwined, which can give rise to complex and collective phenomena. The time-reversal symmetry breaking charge order can open a topological gap on the Dirac nodal lines at the Fermi level, introducing large anomalous Hall effect. It can also be a strong precursor of unconventional superconductivity as we study below. \\



\textbf{Unconventional Superconductivity}\\

The time-reversal symmetry-breaking charge order can arise from extended Coulomb interactions of the kagome lattice with van Hove singularities, where the same interactions and instabilities can lead to correlated superconductivity. Thus, we next focus on the low transverse-field ${\mu}$SR measurements performed in the superconducting state. With a superconducting transition temperature $T_{{\rm c}}$~of~${\simeq}$~1.1~K, the TF-${\mu}$SR spectra above (1.25~K) and below (0.25~K) the superconducting transition temperature $T_{{\rm c}}$ are shown in Fig.~3$\bf{a}$. In order to obtain well ordered vortex lattice, the measurements were done after field cooling the sample from above $T_{\rm c}$.
Above $T_{{\rm c}}$, the oscillations show a damping essentially due to the random local fields from the nuclear magnetic moments. The damping rate is shown to be nearly constant between 10~K and 1.25~K. Below $T_{{\rm c}}$ the damping rate increases with decreasing temperature due to the presence of a nonuniform local magnetic field distribution as a result of the formation of a flux-line lattice in the superconducting state. Figure~3$\bf{b}$ depicts the temperature evolution of the total gaussian relaxation rate  $\sigma$$_{\rm total}$ = $\sqrt{\sigma_{\rm SC}^{2} + \sigma_{\rm nm}^{2}}$  for KV$_{3}$Sb$_{5}$ for the 5~mT and 10~mT fields applied both within and out-of the kagome plane. In order to extract the $\sigma$$_{\rm SC}$ contribution due only to superconductivity, the average value of the normal state depolarization rate $\sigma$$_{\rm nm}$ estimated from six temperature points just above the onset of the superconducting transition has been quadratically subtracted, since above $T_{\rm c}$ there is only the normal state contribution to $\sigma$$_{\rm total}$. Figure 3$\bf{c}$ shows the temperature dependences of the superconducting relaxation rates $\sigma_{\rm SC,ab}$ and $\sigma_{\rm SC,ac}$, determined from the data with the field applied along the $c$-axis and within the kagome plane, respectively. The $c$-axis relaxation rate can be extracted as $\sigma_{\rm SC,c}$ = $\sigma_{\rm SC,ac}^{2}$/$\sigma_{\rm SC,ab}$ \cite{Khasanov}, which is shown as a function of temperature in Fig.~3$\bf{c}$.
 
  We note that the magnetic penetration depth ${\lambda}(T)$ (right axis of Fig. 3$\bf{c}$) is related to the relaxation rate 
${\sigma}_{{\rm SC}}(T)$ in the presence of a triangular (or hexagonal) vortex lattice by the equation \cite{Sonier}: 
\begin{equation}
\frac{\sigma_{SC}(T)}{\gamma_{\mu}}=0.06091\frac{\Phi_{0}}{\lambda^{2}(T)},
\end{equation}
where ${\gamma_{\mu}}$ is the gyromagnetic ratio of the muon and ${\Phi}_{{\rm 0}}$ is the magnetic-flux quantum. 
Since the applied field is a factor of 20-60 times smaller than the second critical magnetic fields (${\mu}_{0}$$H_{\rm c2,c}$ ${\simeq}$ 0.1 T for $H$ ${\parallel}$ $c$ and ${\mu}_{0}$$H_{\rm c2,ab}$ ${\simeq}$ 0.3 T for $H$ ${\parallel}$ $ab$) in KV$_3$Sb$_5$, the Eq.~2 is valid to estimate both the $\lambda_{ab}$ and $\lambda_{c}$. The value of the in-plane penetration depth $\lambda_{ab}$ at 0.3~K, determined from $\sigma_{\rm SC,ab}$ (superconducting
screening currents flowing parallel to the kagome plane), is found to be $\lambda_{ab}$ ${\simeq}$ 877(20)~nm. The value of the out-of-plane penetration depth, determined from $\sigma_{\rm SC,c}$ (superconducting screening currents flowing perpendicular to the kagome plane), is found to be $\lambda_{c}$ ${\simeq}$ 730(20)~nm. The $\lambda(T)$ in the applied field of 5~mT shows a well pronounced two step behavior,  which is reminiscent of the behavior observed in well-known two-band
superconductors with single $T_{\rm c}$ such as FeSe$_{0.94}$ \cite{Khasanov} and V$_{3}$Si \cite{Kogan}. These results were
explained \cite{Kogan} by two nearly decoupled bands with an extremely weak interband coupling (still sufficient to give a single $T_{\rm c}$). According to our numerical analysis (see the methods section) our observation of two step behavior of $\lambda(T)$ in KV$_{3}$Sb$_{5}$ is consistent with two gap superconductivity with very weak interband coupling (0.001-0.005) and strong electron-phonon coupling. The multi gap superconductivity was also recently reported for the sister compound CsV$_{3}$Sb$_{5}$ by means of ${\mu}$SR \cite{GuptaCs} and STM \cite{HShu}. The multi-gap superconductivity in KV$_{3}$Sb$_{5}$ is consistent with the presence of multiple Fermi surfaces revealed by electronic structure calculations and tunneling measurements \cite{Wang2021}.

 To place the system KV$_{3}$Sb$_{5}$  in the context of other superconductors, in Fig.~3$\bf{d}$ we plot the critical temperature $T_{\rm c}$ against the superfluid density $\lambda_{\rm ab}^{-2}$. Most unconventional superconductors have $T_{\rm c}$/$\lambda_{\rm ab}^{-2}$ values of about 0.1$-$20, whereas all of the conventional Bardeen-Cooper-Schrieffer (BCS) superconductors lie on the far right in the plot, with much smaller ratios \cite{Uemura1}. In other words, unconventional superconductors are characterized by a dilute superfluid (low density of Cooper pairs) while conventional BCS superconductors exhibit a dense superfluid. Moreover, a linear relationship between $T_{\rm c}$ and $\lambda_{\rm ab}^{-2}$ is expected only on the Bose Einstein Condensate (BEC)-like side of the phase diagram and is considered a hallmark feature of unconventional superconductivity \cite{Uemura1}, where (on-site or extended) Coulomb interactions plays a role. For KV$_{3}$Sb$_{5}$, the ratio is estimated to be $T_{\rm c}$/$\lambda_{\rm ab}^{-2}$ ${\simeq}$ 0.7, which is far away from conventional BCS superconductors and approximately a factor of two greater than that of charge density wave superconductors 2H-NbSe$_{2}$ and 4H-NbSe$_{2}$ as well as Weyl-superconductor $T_{d}$-MoTe$_{2}$ \cite{GuguchiaNbSe2} and kagome superconductor LaRu$_{3}$Si$_{2}$, as shown in Fig.~3$\bf{d}$. The point for KV$_{3}$Sb$_{5}$ is close to the electron-doped cuprates, which are well-known correlated superconductors with poorly screened Coulomb interactions. \\
  
\textbf{Conclusion}  \\
    
 Our work points to a time-reversal symmetry-breaking charge order, intertwined with correlated superconductivity in the kagome superconductor KV$_{3}$Sb$_{5}$. While low-temperature time-reversal symmetry-breaking superconductivity has been discussed for many systems, high-temperature time-reversal symmetry-breaking charge order is extremely rare, and finds a direct comparison with the fundamental Haldane and Varma models. The complex intertwining of such a charge ordered state with correlated superconductivity highlights the rich nature of the correlated kagome lattice and hints at other hitherto unknown hybrid phenomena resulting from 
nontrivial quantum interactions.

\newpage

\section{METHODS}

\textbf{Sample preparation}: Single crystals of KV$_{3}$Sb$_{5}$ were grown from KSb$_{2}$ alloy as a flux. K, V, Sb elements and KSb$_{2}$ precursor were sealed in a Ta crucible in a molar ratio of 1:3:14:10, which was finally sealed in a highly evacuated quartz tube. The tube was heated up to 1273~K, dwelt for 20 hours and then slowly cooled down to 773~K. Single crystals were separated from the flux by centrifuging. Crystals obtained from flux are thin, metallic platelets with high luster and the largest size of approximately 5~mm~$\times$~5~mm. The obtained crystals with natural hexagonal facets were easily exfoliated. The XRD pattern of a single crystal of KV$_{3}$Sb$_{5}$ was collected using a Bruker D2 Phaser X-ray diffractometer with Cu~$K_{\alpha}$ radiation ($\lambda$~=~0.15418~nm) at room temperature. The single-crystal diffraction was implemented on a Bruker D8 Venture system equipped with a Mo~$K_{\alpha}$ ($\lambda$~=~0.71073~\AA). The crystal structure was solved and refined using the Bruker SHELXTL Software Package. Critical magnetic fields for KV$_{3}$Sb$_{5}$ for the field applied along the $c$-axis and within the kagome plane are 100~mT and 300~mT, respectively.\\

\textbf{Experimental details}: Zero field (ZF) and transverse field (TF) $\mu$SR experiments on the single crystalline samples of KV$_{3}$Sb$_{5}$ were performed on the GPS, Dolly, and high-field HAL-9500 instruments at the Swiss Muon Source (S$\mu$S) at the Paul Scherrer Institut, in Villigen, Switzerland. Zero field is dynamically obtained (compensation better than 30 mG) by a newly installed automatic compensation device \cite{Amato}. When performing measurements in zero-field the geomagnetic field or any stray fields are tabulated and automatically compensated by the automatic compensation device.\\

Because the KV$_{3}$Sb$_{5}$ samples were rather thin (0.13 - 0.35~mm along $c$-axis), a mosaic of several crystals stacked on top of each other was used for these measurements. The individual crystals were attached to a 25~$\mu$m thick Cu foil mounted on the Cu sample fork, and the entire ensemble was held together by small droplets of GE varnish. The crystals were aligned with the same in-plane orientation, which was achievable because the thin sheet-like crystals grow with a clearly hexagonal shape (see Extended data Figure~4c). These multilayer crystal mosaics were then wrapped in a single layer of 60~$\mu$m PE polyester tape. The magnetic field was applied both in-plane (along the $ab$-plane) and out-of-plane (along the crystallographic $c$-axis). A schematic overview of the experimental setup for zero-field and low transverse field measurements in GPS is shown in Figure~1a. The muon spin is forming $60^{\circ}$ with respect to the $c$-axis of the crystal. The sample was surrounded by four detectors: Forward (1), Backward (2), Up (3), and Down (4). A schematic overview of the experimental setup for the high-field $\mu$SR instrument is shown in Figure~2a. The crystals were mounted on 10~mm circular silver sample holder by small droplets of GE varnish. The muon spin forms $90^{\circ}$ with respect to the $c$-axis of the crystal. The sample was surrounded by 2~$\times$~8 positron detectors, arranged in rings. The specimen was mounted in a He gas-flow cryostat with the largest face perpendicular to the muon beam direction, along which the external field was applied.\\

\textbf{${\mu}$SR experiment}: In a ${\mu}$SR experiment nearly 100${\%}$ spin-polarized muons ${\mu}$$^{+}$ are implanted into the sample one at a time. The positively charged ${\mu}$$^{+}$ thermalize at interstitial lattice sites, where they act as magnetic microprobes. In a magnetic material the muon spin precesses in the local field $B_{\rm int}$ at the muon site with the Larmor frequency ${\nu}_{\rm int}$~=~$\gamma_{\rm \mu}$/(2${\pi})$$B_{\rm int}$ (muon gyromagnetic ratio $\gamma_{\rm \mu}$/(2${\pi}$)~=~135.5~MHz~T$^{-1}$). 
Using the $\mu$SR technique, important length scales of superconductors can be measured, namely the magnetic penetration depth $\lambda$ and the coherence length $\xi$. If a type II superconductor is cooled below $T_{\rm c}$ in an applied magnetic field ranging between the lower ($H_{c1}$) and the upper ($H_{c2}$) critical fields, a vortex lattice is formed which in general is incommensurate with the crystal lattice, with vortex cores separated by much larger distances than those of the crystallographic unit cell. Because the implanted muons stop at given crystallographic sites, they will randomly probe the field distribution of the vortex lattice. Such measurements need to be performed in a field applied perpendicular to the initial muon spin polarization (so-called TF configuration). $\lambda$ is one of the fundamental parameters of a superconductor, since it is related to the superfluid density $n_{s}$ via 1/${\lambda}^{2}$ = $\mu_{0}$$e^{2}$$n_{s}/m^{*}$ (where $m^{*}$ is the effective mass).\\
  
\textbf{Analysis of TF-${\mu}$SR data}: The TF ${\mu}$SR data were analyzed by using the following functional form \cite{Bastian}:
\begin{equation}
\begin{aligned}
P_{TF}(t)=A_s\exp\Big[-\frac{(\sigma_{SC}^2+\sigma_{nm}^2)t^2}{2}\Big]\cos(\gamma_{\mu}B_{int,s}t+\varphi) \\
\end{aligned}
\end{equation}
 Here $A_{\rm s}$ denotes the initial asymmetry of the sample. ${\varphi}$ is the initial phase of the muon-spin ensemble and $B_{\rm int}$ represents the internal magnetic field at the muon site. The relaxation rates ${\sigma}_{\rm SC}$ 
and ${\sigma}_{\rm nm}$ characterize the damping due to the formation of the flux-line lattice in the superconducting state and of the nuclear magnetic dipolar contribution, respectively. As indicated by the solid line in Fig.~3a the ${\mu}$SR data are well described by Eq.~(1).\\

\textbf{Crystal structure of KV$_{3}$Sb$_{5}$}: Additional characterization information is provided here on the kagome superconductor KV$_{3}$Sb$_{5}$ which crystallizes in the novel $A$V$_{3}$Sb$_{5}$-type structure (space group $P6$/$mmm$, where $A$ = K,~Rb,~Cs). The crystallographic structure of prototype compound KV$_{3}$Sb$_{5}$ shown in panel (a) of Extended Data Figure 4 illustrates how the V atoms form a kagome lattice (medium beige circles) intertwined with a hexagonal lattice of Sb atoms (small red circles). The K atoms (large purple circles) occupy the interstitial sites between the two parallel kagome planes. In panel (b) the vanadium kagome net has been emphasized, with the interpenetrating antimony lattice included to highlight the unit cell (see dashed lines).
Extended data Figures 4c and d show an optical microscope image of a 3~x~2~x~0.2~mm single crystal of KV$_{3}$Sb$_{5}$ on millimeter paper and Scanning Transmission Microscope (STM) image of the V kagome lattice from a cryogenically cleaved sample, respectively.  \\

\textbf{Extended laboratory X-ray diffraction experiments on a single crystal of KV$_{3}$Sb$_{5}$}: A single crystalline sample has been selected and X-ray diffraction was performed on it. The crystal was oriented such that the incident X-rays scattered off of the $ab$-plane. The resultant diffraction pattern shows clear diffraction peaks, which have been indexed (see the Extended data Figure 5a) and fitted using the SHELX-2018/3 program. The obtained crystallographic information is summarized in Tables 1 and 2. The Laue X-ray diffraction image (see the Extended data Figure 5b) demonstrates the single crystallinity of the samples used for ${\mu}$SR experiments.\\ 

\textbf{Extended magnetic susceptibility measurements of KV$_{3}$Sb$_{5}$}: The magnetic susceptibility measurements show the abrupt drop in susceptibility at $T^{*}$ ${\simeq}$ 80 K (see Extended data Figure 5c) which comes from the charge ordering below this temperature. Such an anomaly in susceptibility across the charge order $T^{*}$ is seen up to the highest magnetic field applied along the $c$-axis. Interestingly, a shallow minimum around 30 K is also seen in macroscopic susceptibility similar to ${\mu}$SR Knight shift data. Whether this increase is related to the appearance of the electronic nematic susceptibility or not is an open question and requires more exploration. Remarkably, the anomaly in magnetic susceptibility across the charge order temperature $T^{*}$ is very well pronounced when the field is applied along the $c$-axis, while it is hardly seen when the field is applied along the kagome plane. This indicates that the magnetic response across $T^{*}$ is anisotropic. We note that field dependence of the charge order peak intensities, observed with STM, is also seen only when the magnetic field is applied along the $c$-axis, pointing to the anisotropic field response of a charge order. These results are consistent with the scenario of orbital currents: since the orbital currents are coupled to the $c$-axis moment and do not produce in-plane fields the pronounced change in susceptibility is seen only when field is applied along the $c$-axis. \\

We note that the temperature dependence of the ${\mu}$SR Knight shift, $K_{exp}$, towards the low temperatures within charge ordered state does not fully coincide with the temperature dependence of the macroscopic susceptibility, e.g., the peak which was seen in $K_{exp}$ near 15 K is missing in the macroscopic susceptibility. We note that in a paramagnetic metal, $K_{exp}$ originates from hyperfine fields produced by the field-induced polarization of conduction electrons and localized electronic moments. The local moments contribute to $K_{exp}$ via two coupling mechanisms: (1) the dipolar interaction between the local moments and the ${\mu}^{+}$, which may be described as a dipolar field at the ${\mu}^{+}$ interstitial site, and (2) a contact term due to electron spin polarization at the interstitial ${\mu}^{+}$. Both contributions are proportional to the local-moment susceptibility. Since there is a possible transition from isotropic charge order to a nematic state in KV$_{3}$Sb$_{5}$ below 30 K, changes in the charge section will modify a hyperfine contact field and thus the local susceptibility. If so, the modified local susceptibility will be reflected in a breakdown of the proportionality of the ${\mu}^{+}$ Knight shift to the measured bulk susceptibility since in this case local susceptibility is different from macroscopic susceptibility. This may explain different temperature dependence of muon Knight shift and macroscopic susceptibility within the charge ordered state.


\section{Crystallographic Structural Information}

$Table$ 1: summary of the atomic positions obtained from fitting to the diffraction data.
\begin{center}
\begin{tabular}{| c | c | c | c | c | c | c |}
\hline
\multicolumn{7}{|c|}{\textbf{Atomic Positions}}\\
\hline \hline
Atom & Wyckhoff Positions & x & y & z & U$_{eq}$/\AA$^{2}$ & Occ\\
\hline
Sb$_{01}$ & 1a & 1 & 1 & 0 & 0.16 &  1\\
Sb$_{02}$ & 4h & 2/3 & 1/3 & 0.25381 & 0.019 & 1\\
V$_{01}$ & 3f & 1/2 & 1/2 & 0 & 0.015 & 1\\
K$_{01}$ & 1b & 1 & 0 & 1/2 & 0.045 & 0.977\\
\hline \hline
\end{tabular}
\end{center}

$Table$ 2: summary of the crystallographic, refinement, and data collection information from the refinements to the single crystal X-ray diffraction data.
\begin{center}
\begin{tabular}{| c | c |}
\hline
\multicolumn{2}{|c|}{\textbf{Crystallographic Data}}\\
\hline \hline
Chemical Formula & K$_{0.98}$V$_{3}$Sb$_{5}$\\
\hline
Formula Weight & 799.81 $\frac{\rm g}{\rm mol}$\\
Crystal System & hexagonal\\
Space Group & $P$6/$mmm$\\
$a$ & 5.4831(3) \AA\\
$c$ & 8.9544(9) \AA\\
$\alpha$ & 90$^{\circ}$\\
$\gamma$ & 120$^{\circ}$\\
$Z$ & 1 \AA\\
Cell Volume & 233.14(3) \AA$^{3}$\\
Density (calculated) & 5.697 $\frac{\rm g}{\rm cm^{3}}$\\
Temperature & 273(2) K\\
Wavelength & Mo $K$$_{\alpha}$ ($\lambda$ = 0.71073 \AA)\\
$\theta$ Range & 2.27$^{\circ}$ to 36.15$^{\circ}$\\
\hline \hline
\multicolumn{2}{|c|}{\textbf{Data Collection}}\\
\hline \hline
Diffractometer & Bruker D8 Venture\\
Index Ranges & -9 $\leq$ $h$ $\leq$ 9\\
      & -9 $\leq$ $k$ $\leq$ 9\\
       & -14 $\leq$ $l$ $\leq$ 14\\
Reflections Collected & 7207\\
Independent Reflections & 278 [$R$$_{int}$ = 0.0380]\\
Absorption Coefficient & 17.469 mm$^{-1}$\\
Structure Solution Tech. & Direct methods\\
Structure Solution Prog. & SHELX-2014/5 (Sheldrick, 2014)\\

\hline \hline
\multicolumn{2}{|c|}{\textbf{Refinement}}\\
\hline \hline
Refinement Method & Full-matrix least-squares on $F$$^{2}$\\
Refinement Program & SHELX-2018/3 (Sheldrick, 2018)\\
Function Minimized & $\Sigma$ $w$($F$$_{o}$$^{2}$+$F$$_{c}$$^{2}$)$^{2}$\\
Data & 278 reflections (271 final)\\
Parameters & 13 parameters\\
Restraints  & 0 restraints\\
Goodness-of-fit on $F$$^{2}$ & 1.233\\
Final $R$ Indices & I $\leq$ $\sigma$(I) $R$$_{1}$=0.0162, $wR$$_{2}$=0.039\\
    & all data; $R$$_{1}$=0.0171, $wR$$_{2}$=0.039\\
Weighting Scheme & $w$ = $\frac{1}{\sigma^{2}(F_{0}^{2})+(0.0130P)^{2}+0.4869P}$\\
     & where $P$ = ($F$$_{o}$$^{2}$+2$F$$_{c}$$^{2}$)/3\\
Extinction Coefficient & 0.0050(10)\\
$\Delta\rho_{\rm Max}$ and $\Delta\rho_{\rm sMin}$ & 1.259 e\AA$^{-3}$ and -0.481 e\AA$^{-3}$\\
R.M.S. Dev. from Mean & 0.164 e\AA$^{3}$\\
\hline
\end{tabular}
\end{center}

\pagebreak

\textbf{Zero-field ${\mu}$SR spectrum for KV$_{3}$Sb$_{5}$}: In the Extended data Figure 7 is displayed the zero-field ${\mu}$SR spectrum recorded at 5 K. The red solid curve 
represents the fit to the recorded time spectra, using only the Gaussian Kubo Toyabe (GKT) function \cite{Toyabe}:

\begin{equation}
\begin{aligned}
P_{ZF}^{GKT}(t) =  \left(\frac{1}{3} + \frac{2}{3}(1 - \Delta^2t^2 ) \exp\Big[-\frac{\Delta^2t^2}{2}\Big]\right) \\ 
\end{aligned}
\end{equation}  
 where ${\Delta}$/${\gamma_{\mu}}$ is the width of the local field distribution due to the nuclear moments and ${\gamma_{\mu}}$/2${\pi}$ = 135.5~MHz/T is the muon gyromagnetic ratio. The Gaussian Kubo-Toyabe depolarization function (Eq. 1) reflects the field distribution at the muon site created by the nuclear moments of the sample. A GKT shape is expected in KV$_{3}$Sb$_{5}$ due to the presence of the dense system of nuclear moments with large values of nuclear spins ($I$ = 3/2 for $^{39}$K, $I$ = 7/2 for $^{51}$V, and $I$ = 5/2 for $^{121}$Sb) in KV$_{3}$Sb$_{5}$ and a high natural abundance. It is clear that the GKT function  alone is not sufficient to fully describe the zero-field ${\mu}$SR spectrum. Multiplying the GKT function by the additional exponential exp(-$\Gamma t$) term (Eq. 2), which is of electronic in origin, is essential to fully describe the spectrum as shown in the Extended data Figure 7a and in the inset. Thus, ZF-${\mu}$SR consists of the muon spin relaxations ${\Delta}$ and ${\Gamma}$ due to nuclear and electronic moments, respectively. This additional exponential rate is higher in the spectra taken from Forward-Backward detectors as shown in the Extended data Figure 7b. 

\begin{equation}
\begin{aligned}
P_{ZF}^{GKT}(t) =  \left(\frac{1}{3} + \frac{2}{3}(1 - \Delta^2t^2 ) \exp\Big[-\frac{\Delta^2t^2}{2}\Big]\right) \exp(-\Gamma t) \\ 
\end{aligned}
\end{equation}  \\

\textbf{Zero-field ${\mu}$SR results for the polycrystalline sample of KV$_{3}$Sb$_{5}$}: We carried out ZF-${\mu}$SR experiments on the polycrystalline sample of KV$_{3}$Sb$_{5}$ and the results are shown in the Extended data Figure 8. The additional exponential term is clearly visible in the spectrum as shown in the Extended data Figure 8a and in the inset, showing the early time behavior. There is a clear increase immediately visible in the exponential relaxation rate upon lowering the temperature below the charge ordering temperature T* (see the Extended data Figure 8b), which is better visible in the Extended data Figure 8c. This observation in the polycrystalline sample of KV$_{3}$Sb$_{5}$ agrees very well with the results on the single crystals. Our results on several KV$_{3}$Sb$_{5}$ samples (both in single crystals and polycrystals) indicate that exponential term is intrinsically present and gets enhanced below $T^{*}$.  \\

\textbf{High-field ${\mu}$SR spectrum for the single crystal of KV$_{3}$Sb$_{5}$ obtained on the instrument HAL-9500}:  During the high-field experiments, a mosaic of several crystals stacked on top of each other was used. The individual crystals were attached to 10 mm circular silver sample holder and the entire ensemble was held together by small droplets of GE varnish. The Extended data Figure 9 shows the probability field distribution, measured at 5 K in the $c$-axis magnetic field of 8 T.
In the whole investigated temperature range, two-component signals were observed: a signal with fast relaxation 0.42 ${\mu}s^{-1}$  (broad signal) and another one with a slow relaxation 0.08 ${\mu}s^{-1}$ (narrow one). The narrow signal arises mostly from the muons stopping in the silver sample holder and its position is a precise measure of the value of the applied magnetic field. The width and the position of the narrow signal is found to be temperature independent (see the inset of the Extended data Figure 9) as expected and thus they were kept constant in the analysis. The relative fraction of the muons stopping in the sample was fixed to the value obtained at the base-$T$ and kept temperature independent. The signal with the fast relaxation, which is shifted towards the lower field from the applied one, arises from the muons stopping in the sample and it takes a major fraction (60-70 ${\%}$) of the ${\mu}$SR signal. This points to the fact that the sample response arises from the bulk of the sample. Based on this two component signal we can determine Knight shift which is defined as $K_{\rm exp}$ = ($B_{\rm int}$ $-$ $B_{\rm ext}$)/$B_{\rm ext}$, where $B_{\rm int}$ and $B_{\rm ext}$ are the internal and externally applied magnetic fields, respectively.\\

\textbf{Superconducting gap symmetry}: Theoretically, several scenarios for electronically mediated, unconventional superconductivity have been discussed \cite{XWu}. In the band structure of the AV$_3$Sb$_5$ materials, van Hove singularities are found close to the Fermi energy - an electronic structural motif shared with other systems such as the cuprate superconductors or Sr$_2$RuO$_4$. As a particular feature of the kagome lattice, however, there is a sublattice interference mechanism \cite{Kiesel} by which the Bloch states near each van Hove point are supported on a distinct sublattice. This promotes the relevance of longer-range interactions and unconventional pairing states. \\
 
 The Extended data Figure 10a shows the temperature dependences of the superconducting relaxation rates $\sigma_{\rm SC,ab}$ and $\sigma_{\rm SC,ac}$, determined from the data with the field applied along the $c$-axis and within the kagome plane, respectively. When a magnetic field is applied along the $c$-axis, superconducting screening currents will flow in the $ab$ plane; whereas in the case of a magnetic field applied in-plane, the superconducting screening currents will flow along the $c$ and $a$ axes. Thus, $\sigma_{\rm SC,ac}$ consists of both the in-plane and $c$-axis contributions. A two-step nature of the superconducting state with the onset of $T{\rm c}$~${\simeq}$~1.1~K is clearly visible in the 5~mT data for both orientations of the magnetic field with respect to the $c$-axis. \\
 
It is worthwhile noticing that the step feature in the temperature dependence of the penetration depth which we observed for KV$_{3}$Sb$_{5}$ is similar to the sudden decrease of the square root of the second moment of the field distribution at the vortex melting temperature in the cuprate high temperature superconductor Bi$_{2.15}$Sr$_{1.85}$CaCu$_{2}$O$_{8+{\Delta}}$ (BSCCO) \cite{Lee}. This process is thermally activated and caused by increased vortex mobility via a loosening of the inter- or intraplanar FLL correlations. This raises the question whether the two-step transition is related to the vortex lattice melting in KV$_{3}$Sb$_{5}$. We note several arguments against such a scenario: \textbf{(1)} In BSCCO, the step feature occurs not in low fields (10 mT, 20 mT) but only in higher fields at which vortex lattice melting takes place \cite{Lee}. In low fields the effects of the thermal fluctuations of the vortex positions on the ${\mu}$SR linewidth are becoming negligible and smooth temperature dependence of the linewidth is observed all the way up to $T_{\rm c}$ \cite{Lee}. In the case of KV$_{3}$Sb$_{5}$  the step like feature is very well pronounced in 5 mT. With the application of 10 mT, the two-step transition becomes somewhat smoothed out and less pronounced. This is in contrast to what we expect within the scenario of vortex lattice melting. \textbf{(2)} The effect of the vortex lattice melting on the ${\mu}$SR lineshape is to change its skewness from positive (ideal static lattice) to a negative value. Thus, vortex lattice melting is clearly reflected in the line shapes. In the case of KV$_{3}$Sb$_{5}$, the SC relaxation rate (${\mu}$SR linewidth) is small due to the long penetration depth and the ${\mu}$SR line is described by symmetric Gaussian line. Thus, it is difficult to check for the vortex lattice melting based on the shape of the field distribution. However, we carried out such an analysis for the sister compound CsV$_{3}$Sb$_{5}$ \cite{GuptaCs} which exhibits higher superconducting critical temperature as well as higher width of the ${\mu}$SR line than the KV$_{3}$Sb$_{5}$ sample. This allows to describe the lineshape more precisely. By analyzing the asymmetric lineshape of the field distribution and skewness parameter as a function of temperature we showed that the FLL in CsV$_{3}$Sb$_{5}$  is well arranged in the superconducting state and it gets slightly distorted only in the vicinity of $T_{\rm c}$. But, no indication of vortex lattice melting was found in CsV$_{3}$Sb$_{5}$ \cite{GuptaCs}. \textbf{(3)} We also note that the superconductors $A$V$_{3}$Sb$_{5}$ are characterized by small superconducting anisotropy.  Anisotropy of penetration depth for CsV$_{3}$Sb$_{5}$ is  2-3, which is two orders of magnitude smaller than the one  for BSCCO. It is rather close to the values reported for Fe-based high temperature superconductors, where no vortex-lattice melting transition is observed. Considering the above arguments, we think that two step temperature dependence of the low field magnetic penetration depth KV$_{3}$Sb$_{5}$ is indeed due to multi-gap superconductivity with extremely small interband coupling \cite{Kogan}. Smearing out the step like feature with increasing the magnetic field may be understood by the tendency of magnetic field towards the suppression of one SC gap or by enhancing interband coupling with higher fields. \\
 
 The temperature dependence of ${\lambda}$ is sensitive to the topology of the SC gap: while in a fully gapped superconductor, $\Delta\lambda^{-2}\left(T\right)\equiv\lambda^{-2}\left(0\right)-\lambda^{-2}\left(T\right)$ vanishes exponentially at low $T$, in a nodal SC it vanishes as a power of $T$. To quantitatively analyze the temperature dependence of the penetration depth ${\lambda}(T)$ \cite{Bastian,GuguchiaMoTe2} for KV$_{3}$Sb$_{5}$ we employ the empirical ${\alpha}$-model. The $\lambda(T)$ in the applied field of 5~mT shows a well pronounced two step behavior, as shown in the Extended data Figure~10a. This suggests that at least two bands are involved in superconductivity and that the interband coupling is extremely small which is sufficient to have same values of $T_{{\rm c}}$ for different bands but still shows the two step temperature behaviour of the penetration depth \cite{Kogan}. With the application of 10~mT, the two-step transition becomes somewhat smoothed out and less pronounced. The results of the  analysis for the ${\mu}_{0}$$H$~=~10~mT using the ${\alpha}$-model  are presented in the Extended data Figure~10b. We consider two different possibilities for the gap function: either a constant gap, $\Delta_{0,i}=\Delta_{i}$, or an angle-dependent $f$-wave gap of the form $\Delta_{0,i}=\Delta_{i}\cos3\varphi$ which is appropriate for a triangular lattice, where $\varphi$ is the polar angle around the Fermi surface. We note that the $d$-wave form $\Delta_{0,i}=\Delta_{i}\cos2\varphi$ makes no real sense on a lattice with 3-fold rotational symmetry. The analysis certainly rules out a simple 1-gap $s$-wave model as an adequate description of ${\lambda^{-2}}$($T$) for KV$_{3}$Sb$_{5}$. The 2-gap $s$+$s$-wave or 2-gap $s$+$f$-wave models line up with the experimental data very well. Due to the lack of data points below 0.25~K, it is difficult to distinguish between $s$+$s$-wave and $s$+$f$-wave models and to give an estimate for the zero-temperature value of ${\lambda}$. We also observe the diamagnetic shift of about 1~G in the SC phase, as shown in the Extended data Figure~10c, which indicates a $T{\rm c}$ of ${\simeq}$ 1.1~K, and furthermore supports the bulk nature of superconductivity in KV$_{3}$Sb$_{5}$. \\

The value of the in-plane penetration depth $\lambda_{ab}$ at 0.3~K, determined from $\sigma_{\rm SC,ab}$ (superconducting
screening currents flowing parallel to the kagome plane), is found to be $\lambda_{ab}$ ${\simeq}$ 877(20)~nm. The value of the out-of-plane penetration depth, determined from $\sigma_{\rm SC,c}$ (superconducting screening currents flowing perpendicular to the kagome plane), is found to be $\lambda_{c}$ ${\simeq}$ 730(20)~nm. We note that these are the upper limits of $\lambda_{ab}$ and $\lambda_{c}$, since they are obtained by assuming a 100~${\%}$ superconducting volume fraction.
On the other hand vortex lattice disorder increases the measured relaxation rate leading to an underestimation of the penetration depth. The not expected observation of slightly smaller penetration depth $\lambda_{c}$ compared to $\lambda_{ab}$ might therefore originate in a slightly stronger disorder of the vortex lattice when the field is applied within the kagome plane. Another possibility for the similar values of the penetration depths $\lambda_{ab}$ and $\lambda_{c}$ in KV$_{3}$Sb$_{5}$ might be the following: we have recently measured SC anisotropy in the related system CsV$_{3}$Sb$_{5}$ \cite{GuptaCs}. The anisotropy of the magnetic penetration depth is found to be around 
$\gamma_{\lambda}$ ${\simeq}$ 3 which is a factor of  3 smaller than the anisotropy of the second magnetic critical field $\gamma_{Hc2}$ ${\simeq}$ 9 \cite{GuptaCs}.  Such a difference was explained in terms of multiband superconductivity in comparison with well established multigap superconductors (MgB$_{2}$, Fe-based, and etc). For KV$_{3}$Sb$_{5}$, the anisotropy of the second critical field is only $\gamma_{Hc2}$ ${\simeq}$ 3. Assuming that both systems CsV$_{3}$Sb$_{5}$ and KV$_{3}$Sb$_{5}$ have similar SC mechanism, we expect the anisotropy of the penetration depth to be much smaller than 3, which makes this system to exhibit an almost isotropic penetration depth. \\

\textbf{Analysis of the temperature dependence of the penetration depth}: The temperature dependence of ${\lambda}$ is particularly sensitive to the topology of the SC gap: while in a fully gapped superconductor, $\Delta\lambda^{-2}\left(T\right)\equiv\lambda^{-2}\left(0\right)-\lambda^{-2}\left(T\right)$ vanishes exponentially at low $T$, in a nodal superconductor it vanishes as a power of $T$. 
${\lambda}$($T$) was calculated within the local (London) approximation (${\lambda}$ ${\gg}$ ${\xi}$) by the following expression \cite{Brandt}:
\begin{equation}
\frac{\lambda^{-2}(T,\Delta_{0,i})}{\lambda^{-2}(0,\Delta_{0,i})}=
1+\frac{1}{\pi}\int_{0}^{2\pi}\int_{\Delta(_{T,\varphi})}^{\infty}(\frac{\partial f}{\partial E})\frac{EdEd\varphi}{\sqrt{E^2-\Delta_i(T,\varphi)^2}},
\end{equation}
where $f=[1+\exp(E/k_{\rm B}T)]^{-1}$ is the Fermi function, ${\varphi}$ is the angle along the Fermi surface, and ${\Delta}_{i}(T,{\varphi})={\Delta}_{0,i}{\Gamma}(T/T_{\rm c})g({\varphi}$)
(${\Delta}_{0,i}$ is the maximum gap value at $T=0$). 
The temperature dependence of the gap is approximated by the expression 
${\Gamma}(T/T_{\rm c})=\tanh{\{}1.82[1.018(T_{\rm c}/T-1)]^{0.51}{\}}$, 
while $g({\varphi}$) describes 
the angular dependence of the gap and it is replaced by 1 for both an $s$-wave and an $s$+$s$-wave gap,
${\mid}\cos(2{\varphi}){\mid}$ for a $d$-wave gap, and ${\mid}\cos(3{\varphi}){\mid}$ for a $f$-wave gap.\\

\textbf{Analysis of the temperature dependence of the penetration depth with a self-consistent approach for a two-band superconductor}: The $\sigma_{\rm sc}(T)$ data collected in presence of 50 G and 100 G field was analyzed within the framework of quasi-classical Eilenberger weak-coupling formalism, where the temperature dependence of the gaps was obtained by solving self-consistent coupled gap equations rather than using the phenomenological $\alpha$-model \cite{Bouquet_EPL_2001, Prozorov_SST_2006,Khasanov_PRL_2007, Khasanov_PRB_2014}, where the latter considers a similar BCS-type temperature dependence for both gaps. The coupled gap equations as introduced by Kogan $et$ $al.$ are \cite{Kogan_PRB_1981}
\begin{widetext}
\begin{equation}
\begin{split}
\delta_1=&\frac{\Delta_1 }{T_c}\frac{1}{2\pi t} = n_1 \lambda_{11}\delta_1\sum_{n=0}^{\infty}\bigg[S+ln\bigg(\frac{1}{t}\bigg)-\bigg(\frac{1}{n+0.5}-\frac{1}{\sqrt{\delta_{1}^2+(n+0.5)^2}}\bigg)\bigg]\\
&+n_2 \lambda_{12}\delta_2\sum_{n=0}^{\infty}\bigg[S+ln\bigg(\frac{1}{t}\bigg)-\bigg(\frac{1}{n+0.5}-\frac{1}{\sqrt{\delta_{2}^2+(n+0.5)^2}}\bigg)\bigg],
\end{split}
\end{equation}
\begin{equation}
\begin{split}
\delta_2=&\frac{\Delta_2 }{T_c}\frac{1}{2\pi t} = n_1 \lambda_{21}\delta_1\sum_{n=0}^{\infty}\bigg[S+ln\bigg(\frac{1}{t}\bigg)-\bigg(\frac{1}{n+0.5}-\frac{1}{\sqrt{\delta_{1}^2+(n+0.5)^2}}\bigg)\bigg]\\
&+n_2 \lambda_{22}\delta_2\sum_{n=0}^{\infty}\bigg[S+ln\bigg(\frac{1}{t}\bigg)-\bigg(\frac{1}{n+0.5}-\frac{1}{\sqrt{\delta_{2}^2+(n+0.5)^2}}\bigg)\bigg],
\end{split}
\end{equation}
\end{widetext}
where
\begin{equation}
S=\frac{n_1\lambda_{11}+n_2\lambda_{22}-\sqrt{(n_1\lambda_{11}-n_2\lambda_{22})^2+4n_1n_2\lambda_{12}^2}}{2n_1n_2(\lambda_{11}\lambda_{22}-\lambda_{12}^2)},
\end{equation}
and $t=T/T_c$ is the reduced temperature. $n_1$ and $n_2$ represent the partial density of states for the corresponding bands at the Fermi level. $\lambda_{11}$ ($\lambda_{22}$) and $\lambda_{12}$ ($\lambda_{21}$)
are the strengths of intraband and the interband coupling, respectively. The two interband coupling strengths are equal, $i.e.$, $\lambda_{12}$ = $\lambda_{21}$ in the notation of Kogan et al. The temperature variation of $\Delta_1(T)$ and $\Delta_2(T)$ obtained after solving aforementioned coupled gap equations are then used to evaluate the temperature dependence of the normalized inverse-square magnetic penetration depth $\lambda_v^{-2}(T)/\lambda_v^{-2}(0)$, which in turn directly related to the superfluid densities [$\rho_v(T)$] for the two bands, by using the expression
\begin{equation}
\rho_v(T)=\frac{\lambda_v(T)^{-2}}{\lambda_v(0)^{-2}}=\delta_v^2\sum_{n=0}^{\infty}\big[\delta_v^2+(n+0.5)^2 \big]^{-\frac{3}{2}},
\end{equation}
where $v$ = 1, 2 are the band indices.

The total superfluid density is then extracted by using the known temperature variation of $\rho_1(T)$ and $\rho_2(T)$ through the following expression:
\begin{equation}
\rho(T)=\gamma \rho_1(T)+(1-\gamma)\rho_2(T),
\end{equation}
where $\gamma$ is the weighting factor for the contribution of superconducting states with the larger gap $\Delta_1$. A detailed description about the model can be found in Ref. \cite{Gupta}. \\
   
The results of these analyses are shown in the Extended data Figure 11a-c. Our numerical analysis reveals that the two step transition in $\sigma_{\rm sc}(T)$ at 10 mT requires the interband coupling constant to be small, 0.005. For 5 mT data, the interband coupling constant was obtained to be 0.001 which is factor of five smaller than the one estimated for 10 mT. This explains why the step-like feature is smoothed out in 10 mT. The small values of interband coupling constants imply that the band(s), where the large and the small superconducting energy gaps are open, become only weakly coupled. One important point is that if we assume the maximum gap-to-$T_{\rm c}$ ratio to be 3.75 (BCS value), which is a limitation of the model, then one can not reproduce the sharp step-like feature in $\sigma_{\rm sc}(T)$. However, considering this limitation of model, then the data are well explained by a large value of 2${\Delta}$/$k_{\rm B}$$T_{\rm c}$ = 7. Our observation of two step behaviour of penetration depth in the system KV$_{3}$Sb$_{5}$ with single Tc is consistent with two gap superconductivity with very weak interband coupling and large value of 2${\Delta}$/$k_{\rm B}$$T_{\rm c}$ = 7.


\textbf{Competing interests:} All authors declare that they have no competing interests.\\

\textbf{Data Availability}: All relevant data are available from the authors. Alternatively, the data can be accessed through the data base at the following link http://musruser.psi.ch/cgi-bin/SearchDB.cgi.\\


\section{Acknowledgments}~
The ${\mu}$SR experiments were carried out at the Swiss Muon Source (S${\mu}$S) Paul Scherrer Insitute, Villigen, Switzerland. 
The magnetization measurements were carried out on the MPMS device of the Laboratory for Multiscale Materials Experiments, Paul Scherrer Institute, Villigen, Switzerland (SNSF grant no. 206021${\_}$139082). Z.G. acknowledges the useful discussions with Robert Scheuermann, Elvezio Morenzoni and Alex Amato. Z.G., C.M., and D.D. thank Christopher Baines for the technical assistance during DOLLY experiments. M.Z.H. acknowledges visiting scientist support from IQIM at the California Institute of Technology. The theory work at Rice has primarily been supported by the U.S. DOE, BES under Award No. DE-SC0018197 and has also been supported by the Robert A. Welch Foundation Grant No. C-1411 (Q.S.). 
The work at Rice university is also supported by U.S. Department of Energy, BES under Grant No. DE-SC0012311 (P.D.). This work is also supported by Beijing Natural Science Foundation (Grants No. Z180008), the National Key Research and Development Program of China (Grants No. 2017YFA0302900), the National Natural Science Foundation of China (Grants No. U2032204). The work of R.G. was supported by the Swiss National Science Foundation (SNF Grant No. 200021${\_}$175935). 

\textbf{\section{Contributions}}
Z.G. supervised the project. Z.G., Y.-X.J., and M.Z.H. conceived the study. Sample growth and single crystal X-ray diffraction experiments: H. Liu and Y.S.; Magnetization and Laue X-ray diffraction experiments: C.M.III., Z.G., M.M., H.Lei.; ${\mu}$SR experiments and corresponding discussions: Z.G., C.M.III., D.D., R.G., R.K., H.L., J.J.C., J.-X. Yin., Y.-X.J., M.Z.H., X.W., P.D., Q.S., H.M., R.T., T.N.; ${\mu}$SR data analysis: Z.G., and C.M.III., with contributions from H.L., R.K., D.D., R.G.; STM experiments and corresponding discussions: J.-X.Y., 
Y.-X.J., Z.G., and M.Z.H.; Figure development and writing the paper: Z.G.; C.M.III., J.-X.Y., H.L., M.Z.H. with contributions from all authors. All authors discussed the results, interpretation and conclusion.\\

\newpage   

\renewcommand{\figurename}{Extended Data Figure}
   
\begin{figure*}[t!]
\includegraphics[width=1.0\linewidth]{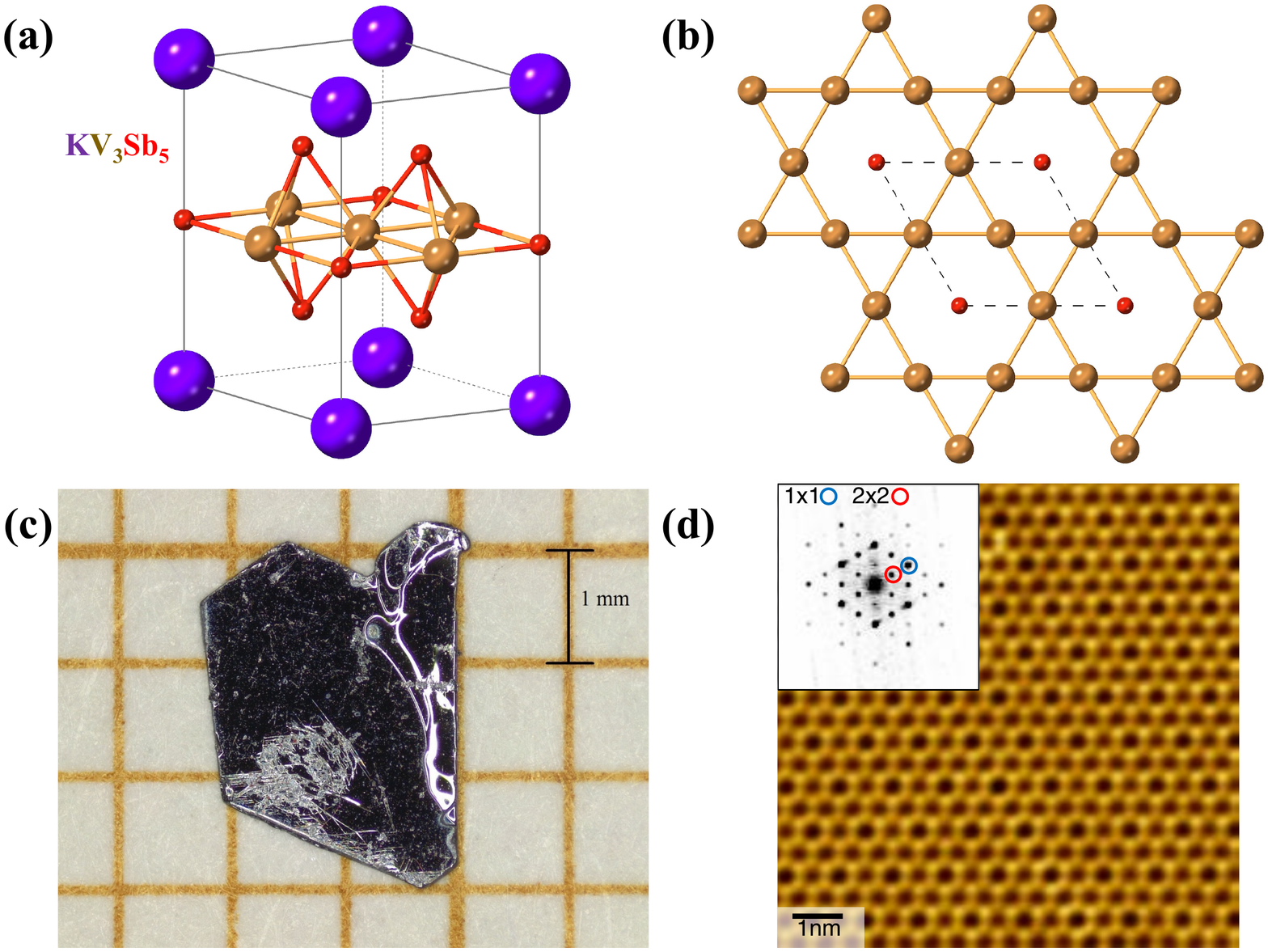}
\vspace{-0.7cm}
\caption{ \textbf{Crystal structure of KV$_{3}$Sb$_{5}$.}
Three dimensional representation (a) and top view (b) of the atomic structure of KV$_{3}$Sb$_{5}$.
In panel (c) is displayed an optical microscope image of a 3~x~2~x~0.2~mm single crystal of KV$_{3}$Sb$_{5}$ on millimeter paper, with the scale shown. The hexagonal symmetry is immediately apparent. (d) Scanning Transmission Microscope (STM) image of the V kagome lattice from a cryogenically cleaved sample.}
\label{fig1}
\end{figure*}

\begin{figure*}[t!]
\includegraphics[width=1.0\linewidth]{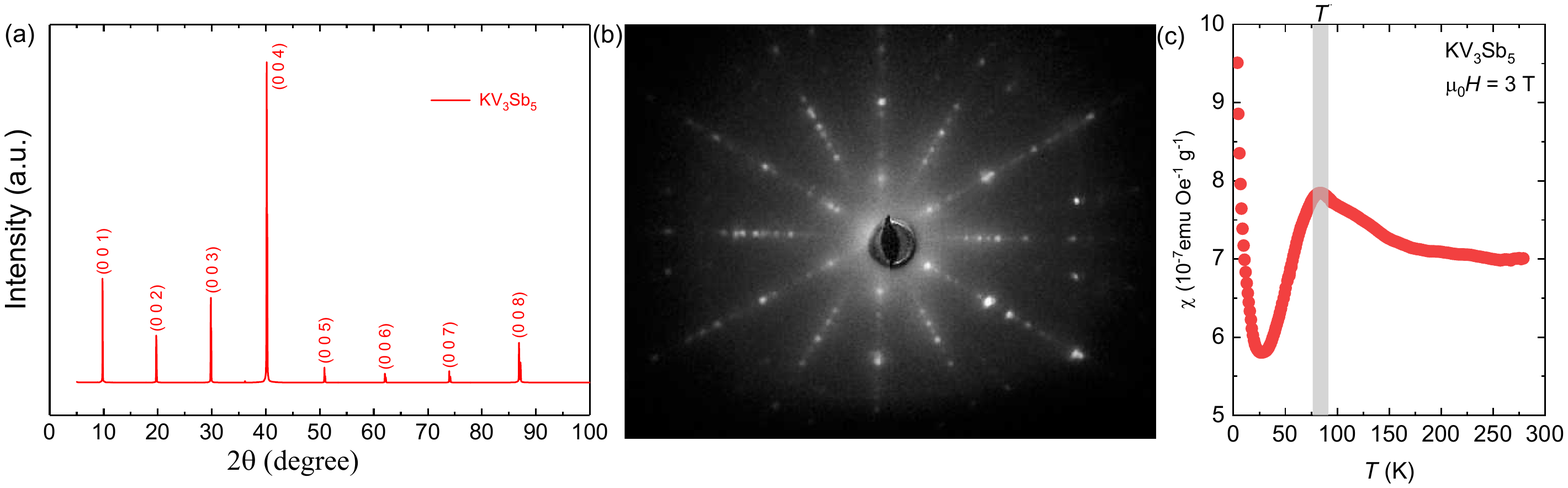}
\vspace{-6.0cm}
\caption{ \textbf{Single Crystal X-Ray Diffraction for KV$_{3}$Sb$_{5}$.}
(a) X-ray diffraction image for KV$_{3}$Sb$_{5}$ recorded at 300 K. The well-defined peaks are labeled with their crystallographic indices. No second phase has been detected. (b) Laue X-ray diffraction image of the single crystal sample KV$_{3}$Sb$_{5}$, oriented with the $c$-axis along the beam. (c) The temperature dependence of magnetic susceptibility of KV$_{3}$Sb$_{5}$ above 1.8 K. It shows an anomaly at $T^{*}$ ${\simeq}$ 80 K, coinciding with emergence of a charge order.}
\label{fig1}
\end{figure*}

\begin{figure*}[t!]
\includegraphics[width=0.7\linewidth]{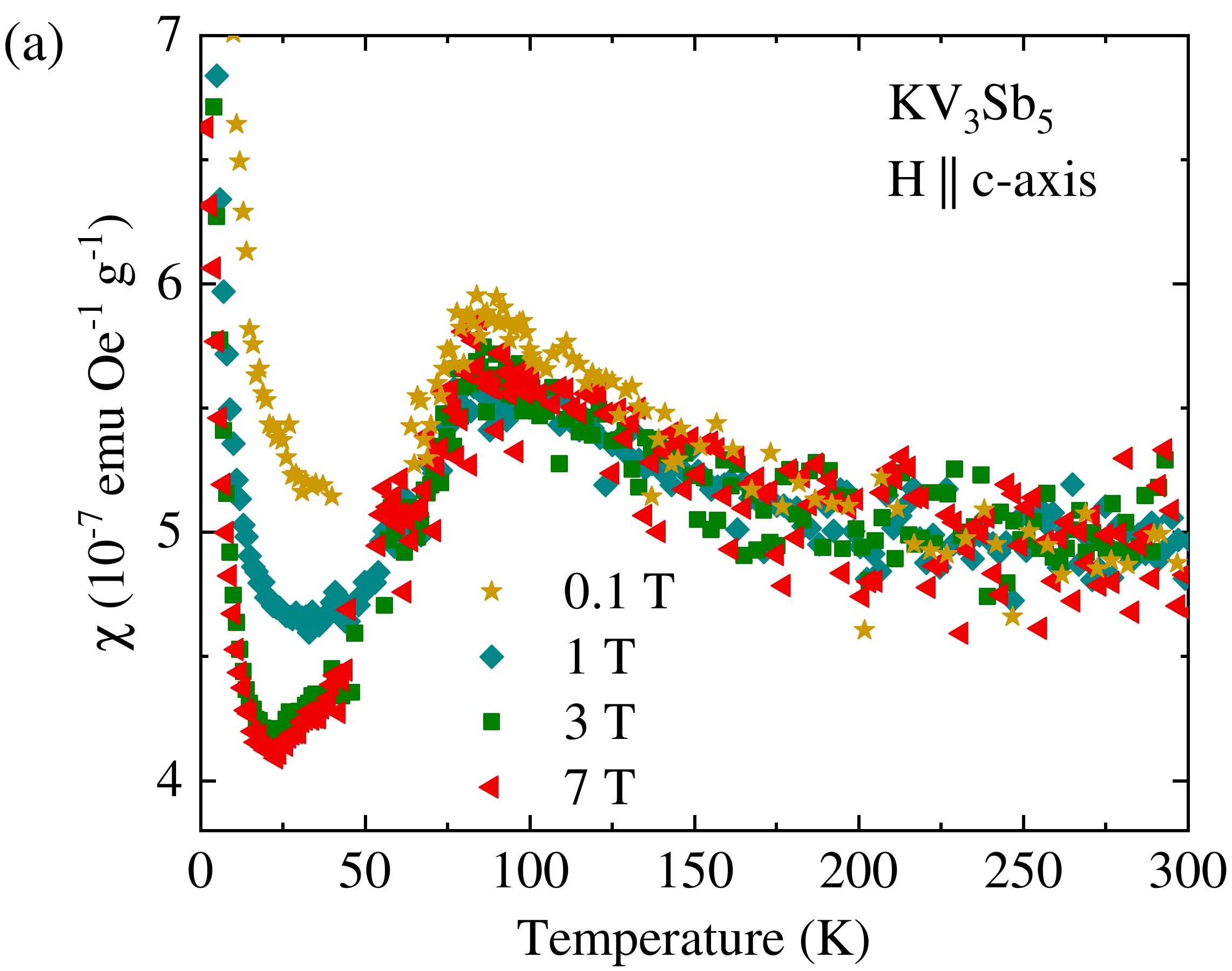}
\includegraphics[width=0.7\linewidth]{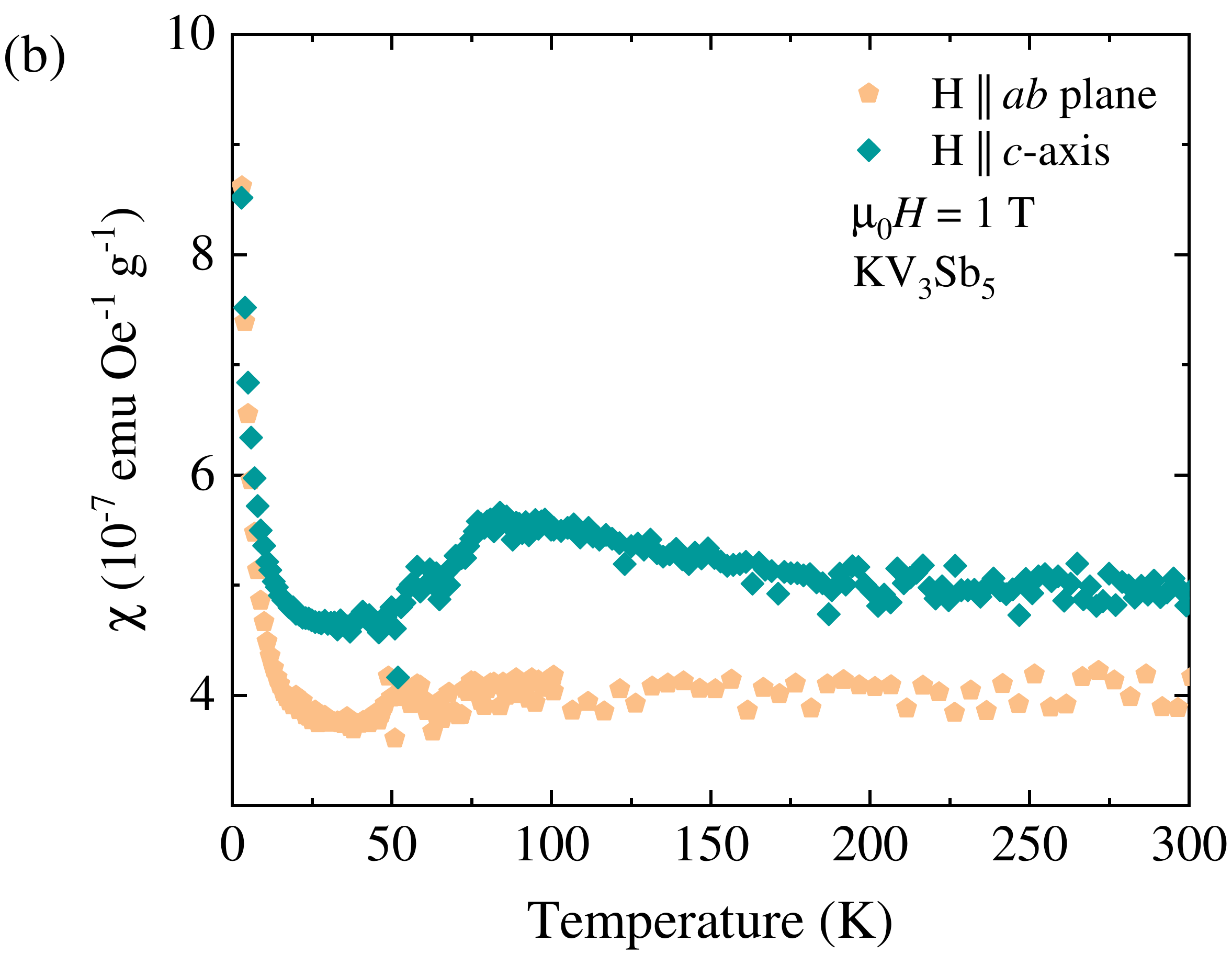}
\vspace{0.0cm}
\caption{ \textbf{Anisotropic magnetic response across charge order temperature in the single crystalline sample of KV$_{3}$Sb$_{5}$.}
(a) The temperature dependence of magnetic susceptibility for KV$_{3}$Sb$_{5}$ measured at various magnetic fields applied parallel to the $c$-axis. (b) The temperature dependence of magnetic susceptibility for KV$_{3}$Sb$_{5}$ measured in the field of 1 T, applied both parallel to the kagome plane and parallel to the $c$-axis.}
\label{fig1}
\end{figure*}

\begin{figure*}[t!]
\includegraphics[width=0.7\linewidth]{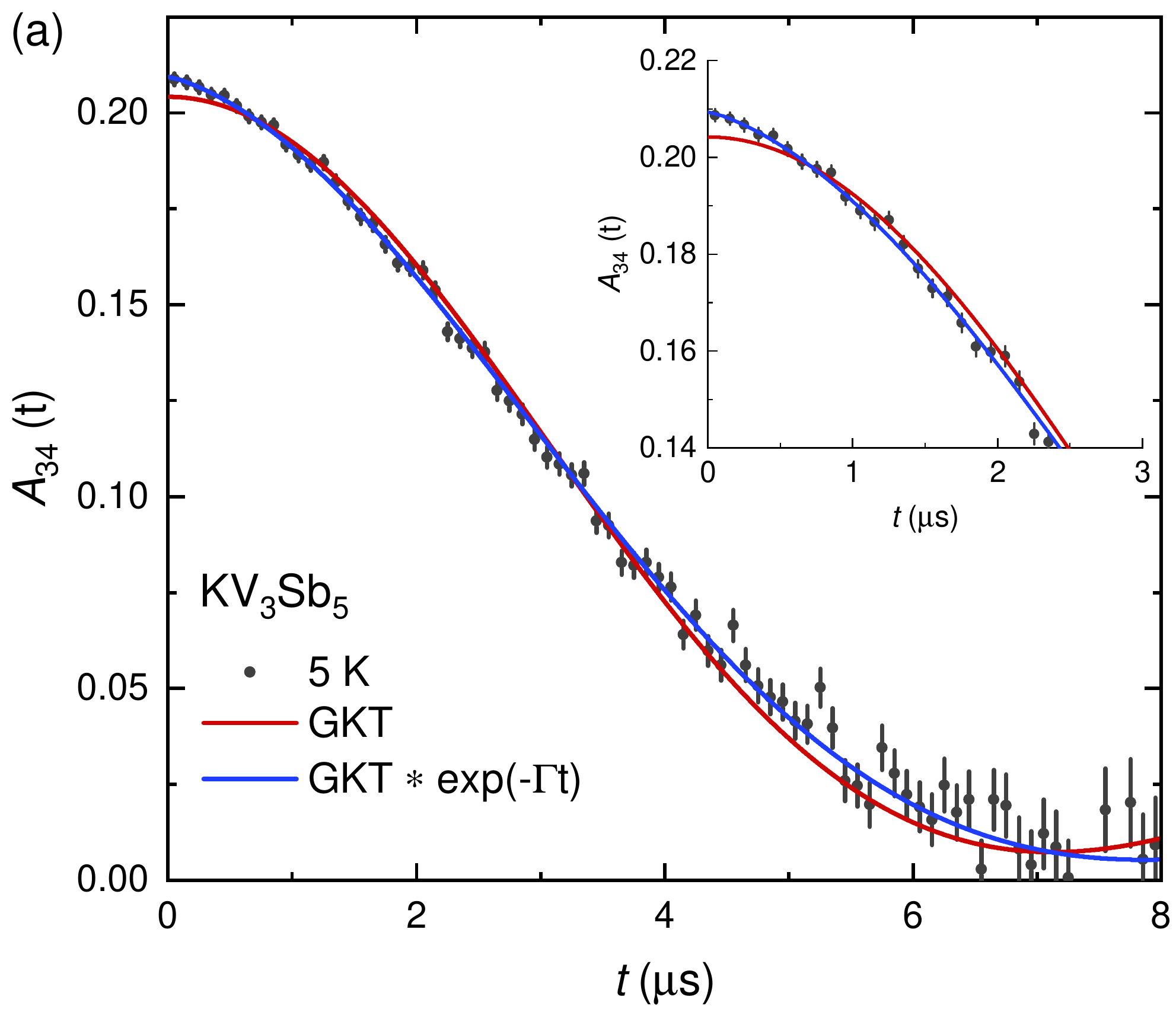}
\includegraphics[width=0.7\linewidth]{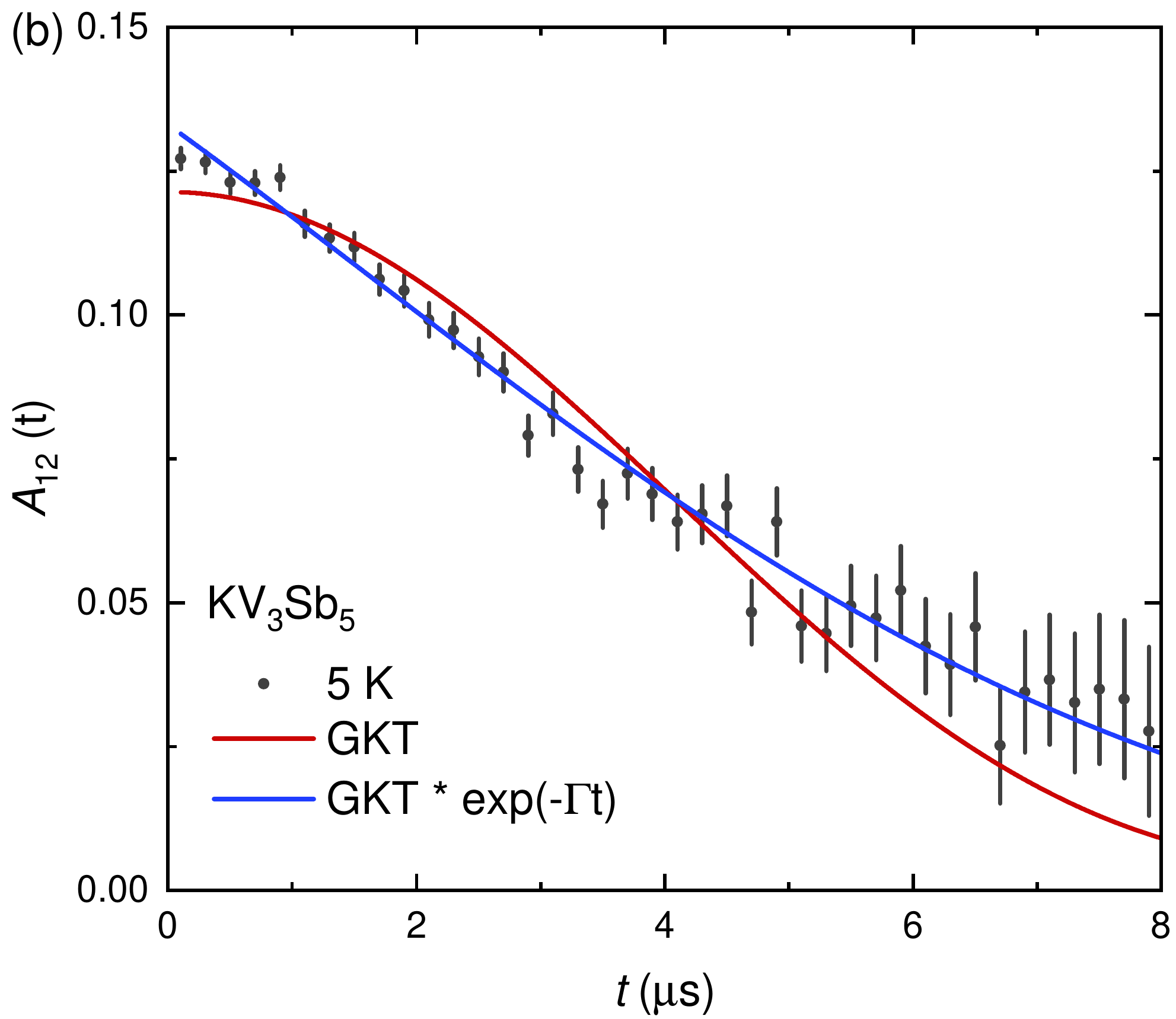}
\vspace{-0.0cm}
\caption{ \textbf{Zero-field ${\mu}$SR experiment for the single crystalline sample of KV$_{3}$Sb$_{5}$.}
The ZF ${\mu}$SR time spectra for KV$_{3}$Sb$_{5}$, obtained at $T$ = 5 K from detectors 3 \& 4 and 2 \& 1. The solid curves represent fits to the recorded time spectra, using only Gaussian Kubo Toyabe (GKT) function (red) and the one with an additional exponential exp(-$\Gamma t$) term (blue). The inset shows the low time part of the spectrum.}
\label{fig1}
\end{figure*}

\begin{figure*}[t!]
\includegraphics[width=0.5\linewidth]{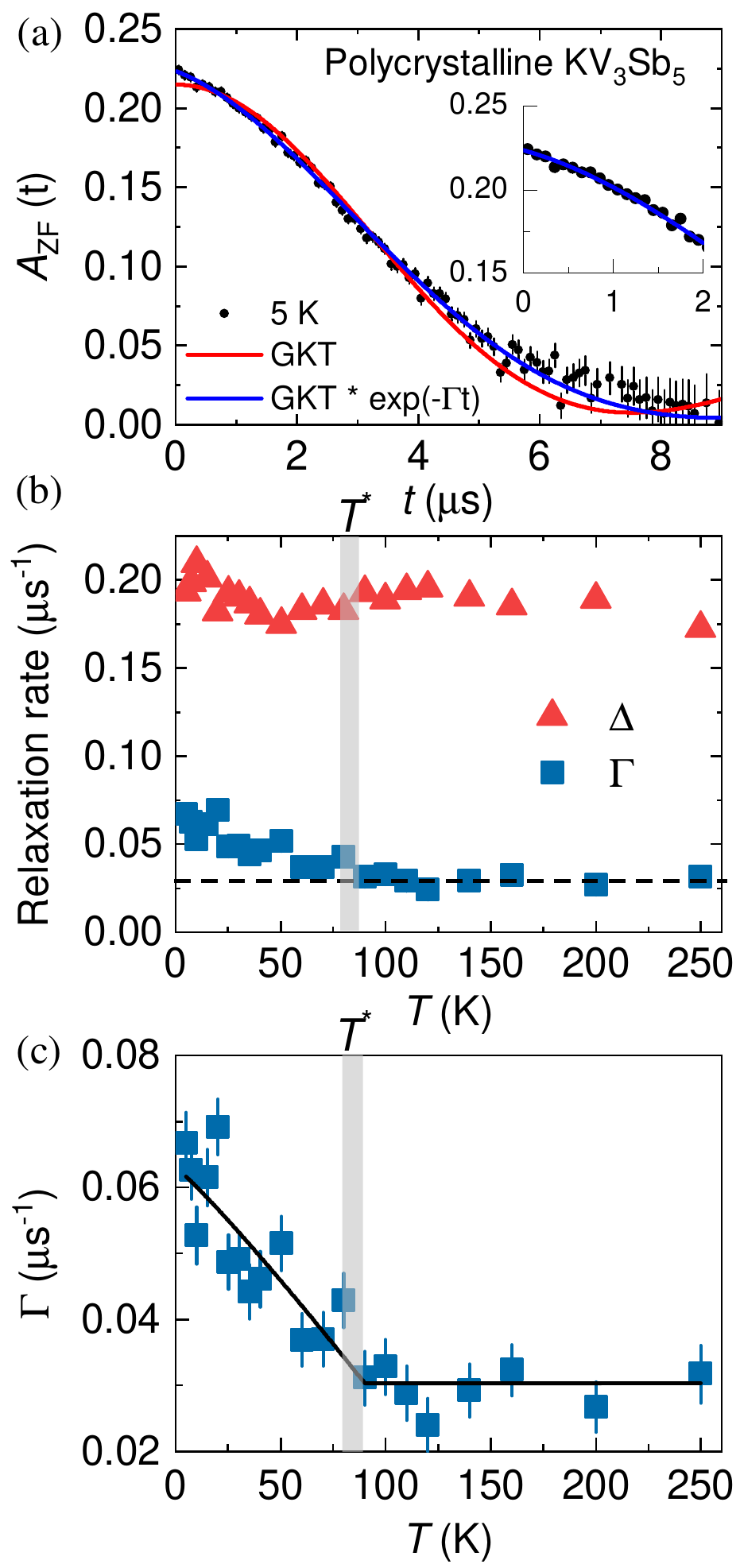}
\vspace{-0.0cm}
\caption{ \textbf{Zero-field ${\mu}$SR experiment for the polycrystalline sample of KV$_{3}$Sb$_{5}$.}
The ZF ${\mu}$SR time spectra for the polycrystalline sample of KV$_{3}$Sb$_{5}$, obtained at $T$ = 5 K. The solid curves represent fits to the recorded time spectra, using only Gaussian Kubo Toyabe (GKT) function (red) and the one with an additional exponential exp(-$\Gamma t$) term (blue). The inset shows the low time part of the spectrum.}
\label{fig1}
\end{figure*}

\begin{figure*}[t!]
\includegraphics[width=1.0\linewidth]{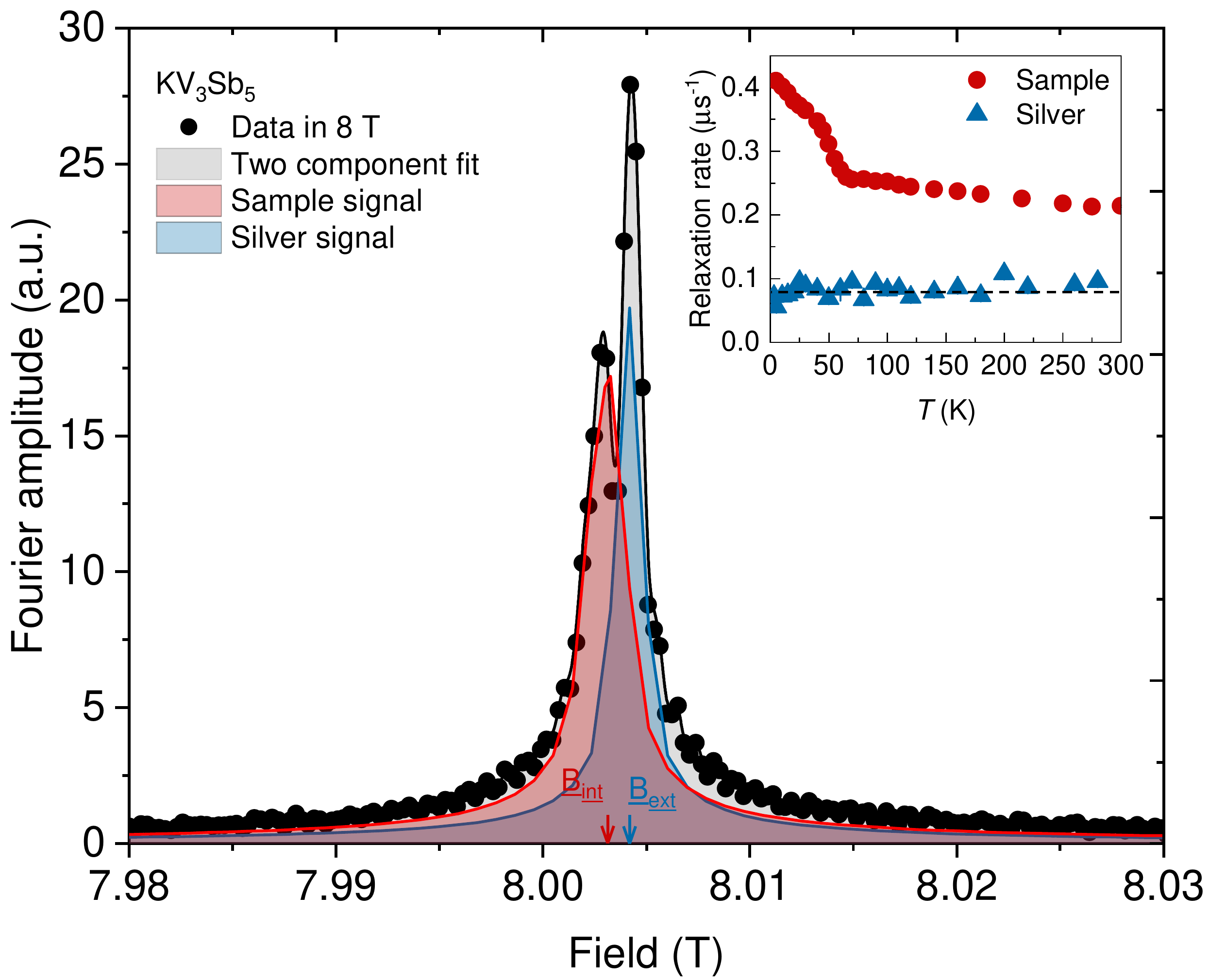}
\vspace{-0.0cm}
\caption{ \textbf{High-field ${\mu}$SR experiment for KV$_{3}$Sb$_{5}$.}
Fourier transform for the ${\mu}$SR asymmetry spectra of KV$_{3}$Sb$_{5}$ at 5 K for the applied field of ${\mu}_{0}$$H$ = 8 T. The black solid line represents the fit to the data using the two component signal. Red and blue solid lines show the signals arising from the sample and the silver sample holder (mostly), respectively. The inset shows the temperature dependences of the muon spin relaxation rates arising from the sample and the silver sample holder.}
\label{fig1}
\end{figure*}

\begin{figure*}[t!]
\includegraphics[width=0.6\linewidth]{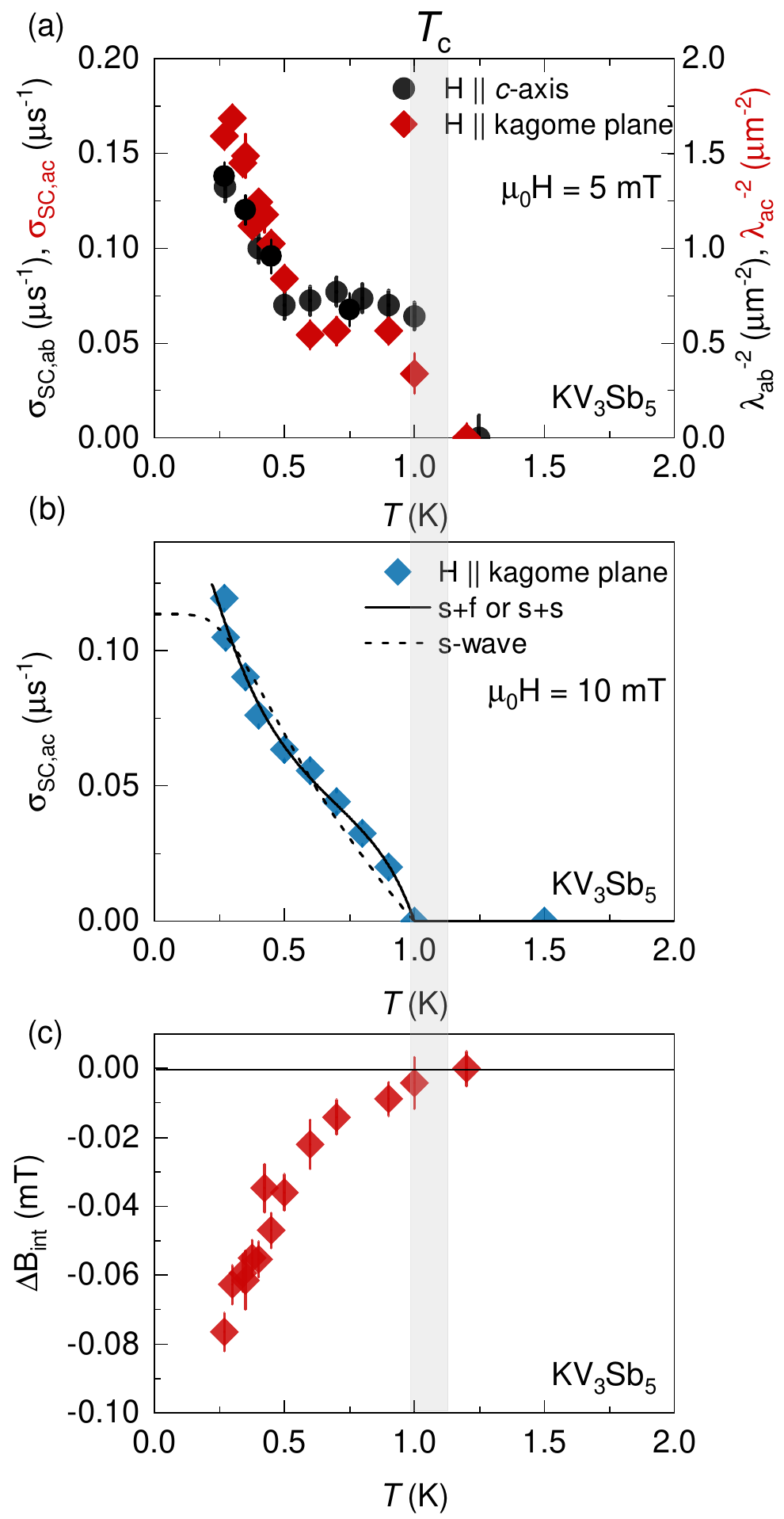}
\vspace{-0.0cm}
\caption{ \textbf{Superconducting gap symmetry in KV$_{3}$Sb$_{5}$.}
(a) The SC muon depolarization rates ${\sigma}_{SC,ab}$, and ${\sigma}_{SC,ac}$ as well as the inverse squared magnetic penetration depth ${\lambda}_{ab}^{-2}$ and ${\lambda}_{ac}^{-2}$ as a function of temperature, measured in 5 mT, applied parallel and perpendicular to the kagome plane. (b) The SC muon depolarization rate ${\sigma}_{SC,ac}$, measured in 10 mT, applied parallel to the kagome plane. The solid line represents the indistinguishable 2-gap $s$-wave and $s+d$ wave model. The error bars represent the s.d. of the fit parameters. (c) Temperature dependence of the difference between the internal field ${\mu}_{\rm 0}$$H_{\rm SC}$ measured in the SC state and the one measured in the normal state ${\mu}_{\rm 0}$$H_{\rm NS}$ at $T$ = 5~K for KV$_{3}$Sb$_{5}$.}
\label{fig1}
\end{figure*}

\begin{figure*}[t!]
\includegraphics[width=0.9\linewidth]{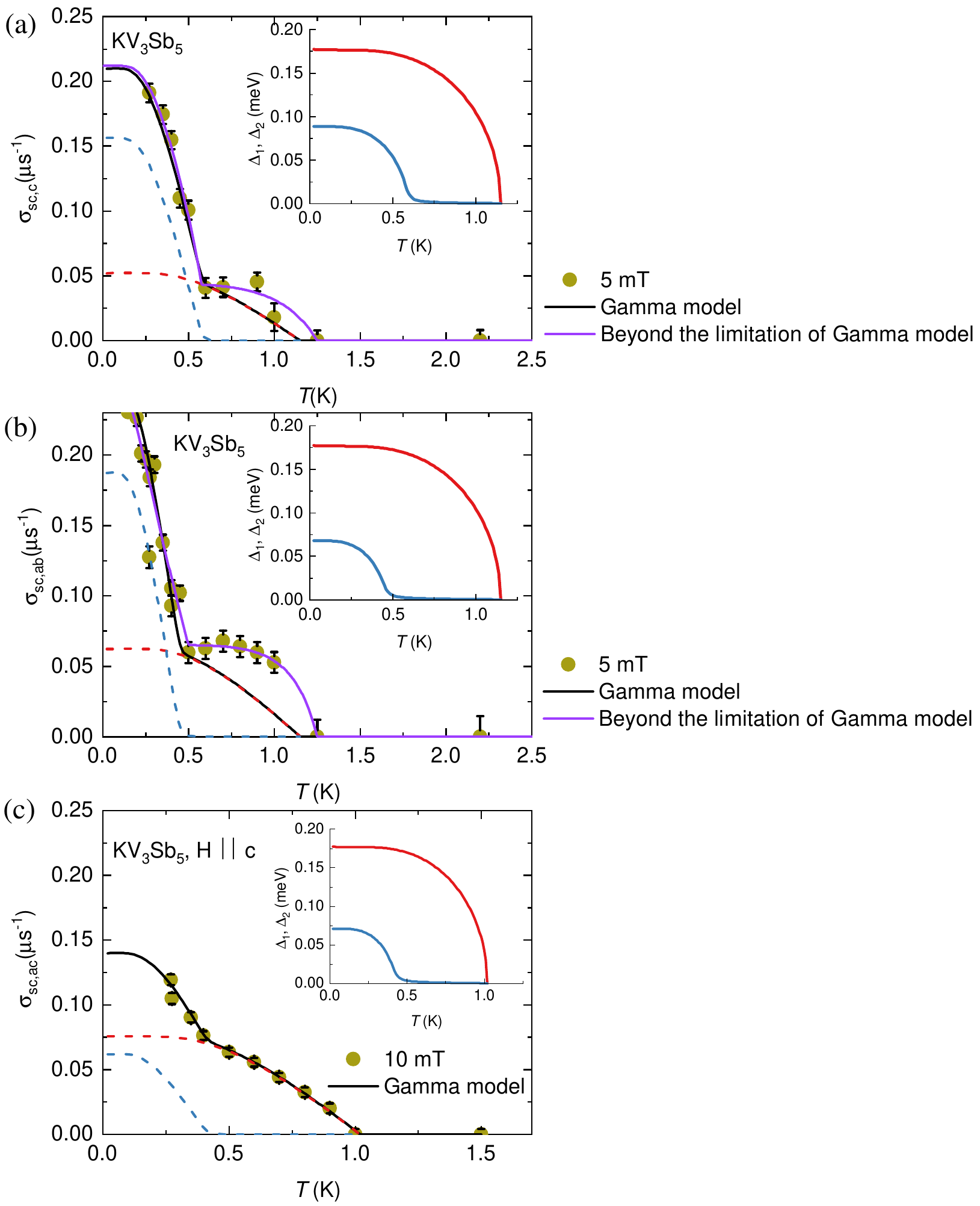}
\vspace{-0.0cm}
\caption{ \textbf{A self-consistent approach for a two-band superconductor in KV$_{3}$Sb$_{5}$.}
The SC muon depolarization rates ${\sigma}_{SC,c}$ (a), and ${\sigma}_{SC,ab}$ (b) as a function of temperature, measured in 5 mT, applied perpendicular and parallel to the kagome plane. (c) The SC muon depolarization rate ${\sigma}_{SC,ac}$, measured in 10 mT, applied parallel to the kagome plane.  The solid black and purple lines are the theoretical curves obtained within the framework of self-consistent approach for a two-band superconductor described in the text. 
The red and the blue dashed lines correspond to the contribution of the large and the small superconducting gaps to the
total superfluid density, solid black lines. The insets show the temperature dependences of the large ${\Delta}_{1}$ and the small ${\Delta}_{2}$.}
\label{fig1}
\end{figure*}

\end{document}